\newcommand{\bB}{\mathbf{B}}
\newcommand{\bv}{\mathbf{v}}
\newcommand{\bA}{\mathbf{A}}
\newcommand{\bJ}{\mathbf{J}}
\newcommand{\nn}{\nonumber}

\newcommand{\nb}{\nabla}

\newcommand{\Ac}{ {\mathcal{A}} }

       \newcommand{\Cc}{ {\mathcal{C}} }

       \newcommand{\Fc}{ {\mathcal{F}} }
       \newcommand{\Gc}{ {\mathcal{G}} }
       \newcommand{\Hc}{ {\mathcal{H}} }

       \newcommand{\Kc}{ {\mathcal{K}} }
       
       \newcommand{\Mc}{ {\mathcal{M}} }
       \newcommand{\Nc}{ {\mathcal{N}} }
       \newcommand{\Oc}{ {\mathcal{O}} }
       
       \newcommand{\Qc}{ {\mathcal{Q}} }

       \newcommand{\R}{ {\mathbb{R}} }
       
       \newcommand{\hz}{ {\hat{z}} }

\documentclass[twocolumn,preprintnumbers,amsmath,amssymb,aps,pop]{revtex4-1}
\usepackage{mathtools}
\usepackage{amsfonts}
\usepackage{graphicx}
\usepackage{dcolumn}
\usepackage{bm}
\usepackage[export]{adjustbox}
\usepackage{wrapfig}
\usepackage{lipsum}
\usepackage[titletoc,toc,title]{appendix}
\usepackage[colorlinks]{hyperref}
\usepackage{color}

\begin{document}

\title{Translationally symmetric extended MHD via Hamiltonian reduction: Energy-Casimir equilibria}

\author{D. A. Kaltsas}
 \email{dkaltsas@cc.uoi.gr}
\author{G. N. Throumoulopoulos}%
 \email{gthroum@cc.uoi.gr}
\affiliation{
Department of Physics, University of Ioannina,\\ GR 451 10 Ioannina, Greece
}%
 \author{P. J. Morrison}
 \email{morrison@physics.utexas.edu}
\affiliation{%
Department of Physics and Institute for Fusion Studies,\\ University of Texas, Austin, Texas 78712, USA 
}

\date{\today}

\begin{abstract}
The Hamiltonian structure of ideal translationally symmetric extended MHD (XMHD) is obtained
by employing a method of Hamiltonian reduction on the three-dimensional noncanonical Poisson
bracket of XMHD. The existence of the continuous spatial translation symmetry allows the introduction of Clebsch-like forms for the magnetic and velocity fields. Upon employing the chain rule for
functional derivatives, the 3D Poisson bracket is reduced to its symmetric counterpart. The sets of
symmetric Hall, Inertial, and extended MHD Casimir invariants are identified, and used to obtain
energy-Casimir variational principles for generalized XMHD equilibrium equations with arbitrary
macroscopic flows. The obtained set of generalized equations is cast into Grad-Shafranov-Bernoulli
(GSB) type, and special cases are investigated: static plasmas, equilibria with longitudinal flows
only, and Hall MHD equilibria, where the electron inertia is neglected. The barotropic Hall MHD
equilibrium equations are derived as a limiting case of the XMHD GSB system, and a numerically
computed equilibrium configuration is presented that shows the separation of ion-flow from electromagnetic surfaces.
\end{abstract}

\pacs{Valid PACS appear here}
\maketitle


\section{Introduction}
\label{sec_I}

By extended MHD (XMHD) we mean  the  one-fluid model  obtained by  reduction of the standard two-fluid plasma model,  when the quasineutrality  assumption  is imposed and expansion in the smallness of the electron mass is performed (e.g.~\cite{lust}).  The resulting  model has  a generalized Ohm's law that contains Hall drift and electron inertia physics, and it  was proven in  \cite{kimu_morr} that  energy conservation for ideal XMHD, the version treated in this paper,  requires the addition of a commonly neglected term in  the momentum equation related to the electron inertia.  

Despite the complexity of XMHD, it was  shown in \cite{abde_kawa_yosh} and \cite{ling_morr_milo} to possess  Hamiltonian structure.  Moreover  in \cite{ling_morr_milo},   remarkable connections with the Hamiltonian structure of other models were  established,  viz., Hall MHD (HMHD) (e.g.\ \cite{lighthill}), Inertial MHD (IMHD) (e.g.\ \cite{kimu_morr,ling_morr_tass}) and usual ideal  MHD (highlighted in   \cite{ling_milo_morr}).  In addition,  the  derivation of XMHD and its Hamiltonian structure from its underlying Lagrangian variable action functional was reported in \cite{avig-morr-ling}.  Recently,  the Hamiltonian structure of two-dimensional incompressible XMHD was derived in \cite{gras_tass_abde_morr}, a reduced XMHD (RXMHD)  that was used to study  Hamiltonian reconnection due to the Hall and electron inertial terms.  The Hamiltonian structure of a similar collisionless fluid reconnection model was established earlier in \cite{tass_morr_wael}, and a general treatment of reduced Hamiltonian models was given in \cite{tass}. 

Detailed consequences of the original noncanonical Hamiltonian structure of  Morrison and Greene \cite{morr_gree}, were explored in a series of papers \cite{andr_morr_pego,amp1,moawad,amp2a,amp2b}, including various variational principles for equilibria and their use in ascertaining stability via energy principles that incorporate different constraints.  Given that XMHD is a Hamiltonian theory and that the investigations of \cite{andr_morr_pego,amp1,moawad,amp2a,amp2b} are generic to Hamiltonian theories,  all of the considerations of these and other works  can be worked out for XMHD. This is the main motivation for conducting this study in the framework of noncanonical Hamiltonian mechanics, i.e.\  since XMHD is a Hamiltonian theory, the existence of the aforementioned variational principles provides us with a joint tool for the derivation of equilibrium equations and stability criteria. This study is focused on equilibria but it may serve as a starting point for a stability analyses as well. Also the Hamiltonian formalism is helpful in order to analyze and describe the geometrical structure of the dynamics, e.g.\  the existence of the so-called Casimir invariants that affect topological structure of the phase-space, by constraining the dynamics to evolve within specific regions. Lastly, the Hamiltonian description may provide the means for the construction of conservative algorithms for numerical analyses
\cite{pjm_str_pres}. All these indicate that a Hamiltonian description, whenever possible, is preferable. 

The present paper considers the case of  translationally symmetric compressible plasmas with an  emphasis on  equilibrium analyses. We derive the Hamiltonian structure of this translationally symmetric model by applying a method of Hamiltonian reduction, which was used  in \cite{andr_morr_pego}, on the parent three-dimensional (3D)  model. Specifically we employ a chain rule reduction on the functional derivatives of  the noncanonical Poisson bracket of XMHD in order to obtain a bracket expressed in terms of Clebsch-like variables that  describe globally the velocity and magnetic field. This reduction leads to the identification of translationally symmetric Casimir invariants,  which due to the spatial symmetry,  form infinite families of generalized helicities. Exploiting these invariants along with the Hamiltonian functional, written in terms of the aforementioned variables, we formulate an Energy-Casimir variational principle that leads to generalized equilibrium equations describing translationally symmetric XMHD equilibria with flows, a set of equations  that we cast into the form of a Grad-Shafranov-Bernoulli system, which makes the equilibrium study of the model tractable.

In comparison to MHD, equilibrium and stability calculations for  XMHD  are considerably more complex.  This is because XMHD contains additional physics, viz.,  XMHD includes the two-fluid phenomena of  Hall drift and electron inertia, arising from the individual fluid dynamics of ions and electrons, while maintaining quasineutrality.  This gives rise to a plethora of new effects as evidenced by the complexity of the linear modes present in XMHD (see e.g.\ Refs.~\onlinecite{lust, kawazura}).  Even in a reduced two-dimensional case,  linear and nonlinear physics is significantly modified and the phenomenon of collisionless reconnection emerges \cite{{gras_tass_abde_morr}}. 
 Even the simpler case of HMHD, which we will address, contains significant complexities not included in  MHD.

The present study is organized as follows: in Sec.~\ref{sec_II} we review the Hamiltonian field theory of XMHD, along with some basic aspects and features of noncanonical Hamiltonian mechanics.  Before proceeding to the Hamiltonian reduction, we present as a preliminary application, the  3D Energy-Casimir variational principle for deriving equilibrium conditions.  In Sec.~\ref{sec_III} we introduce appropriate representations for the magnetic  and  velocity fields, which ensures that they respect translation symmetry and additionally renders the magnetic field   divergence free.  Using this representation,   the Hamiltonian and the XMHD Poisson bracket are reduced to their translationally symmetric counterparts. The Casimir invariants of the symmetric Poisson bracket are computed and their HMHD and IMHD analogues are presented.  In Sec.~\ref{sec_IV} we establish the symmetric variational principle, from which we  derive generalized equilibrium equations. Special cases of equilibria are discussed and studied in detail as applications.  In Sec.~\ref{sec_V} we   conclude  and discuss  extensions of the present study.

\section{Noncanonical Hamiltonian structure of   XMHD}
\label{sec_II}

\subsection{Hamiltonian formulation}

The dynamical equations of the XMHD  model, written in the standard Alfv\'en units,  are the following:
\begin{eqnarray}
\partial_{t}\rho&=&-\nabla\cdot\left(\rho \mathbf{v}\right)\,,
\label{ConDen} \\
\partial_t \bv&=&\bv\times\nabla\times\bv-\nabla\left(h+\frac{v^2}{2}\right)+\rho^{-1}\bJ\times\bB^*
\nonumber\\
&&\hspace{1cm}  -d_e^2\nabla\left(\frac{|\nb\times\bB|^2}{2\rho^2}\right)\,, \\
\partial_t\bB^*&=&\nabla\times\left(\bv\times\bB^*\right)-d_i\nabla\times\left[\frac{\bJ\times \bB^*}{\rho}\right]\nonumber\\
&&\hspace{1cm}  +d_e^2\nabla\times\left[\frac{\bJ\times\left(\nabla\times\bv\right)}{\rho}\right] \,,
\label{starB}
\end{eqnarray}
where
\begin{eqnarray}
\bJ&=&\nabla\times \bB \,,\\
\bB^*&=&\bB+d_e^2\nabla\times\left(\frac{\nb\times\bB}{\rho}\right)\,.
\end{eqnarray}
Here,  a barotropic equation of state has been assumed, which means  the enthalpy $h$ is related to pressure by $\nabla h=\rho^{-1}\nabla p$,  and the parameters $d_i$ and $d_e$ are the normalized ion and electron skin depths, respectively, with $d_s={c}/({\omega_{ps}L})$ and  $s=i,e$. 

As already mentioned in Sec.~\ref{sec_I}, the Hamiltonian structure of the XMHD model of  Eqs.~\eqref{ConDen}--\eqref{starB} was obtained in \cite{abde_kawa_yosh} and \cite{ling_morr_milo}. More precisely it was shown that the equations of motion can be reproduced by using a Hamiltonian ``apparatus" (see  \cite{morr_rev} for a comprehensive review) that  consists of the Hamiltonian function
\begin{eqnarray}
\Hc&=&\frac{1}{2}\int_Vd^3x\,\left[\rho v^2+2\rho U(\rho)+B^2+d_e^2\frac{|\nb\times\bB|^2}{\rho}\right]
\nn\\
&=&\frac{1}{2}\int_V d^3x\,\left[\rho v^2+2\rho U(\rho)+\bB\cdot\bB^*\right]\,,
 \label{hamiltonian}
\end{eqnarray} 
where $V\subseteq \R^3$,  and the  noncanonical Poisson bracket 
\begin{eqnarray}
\left\{F,G\right\}&=&\int_Vd^3x\,\big\{G_\rho\nabla\cdot F_\bv - F_\rho\nabla\cdot G_\bv
\label{poisson}\\
&+&\rho^{-1}\left(\nabla\times\bv\right)\cdot\left(F_\bv \times G_\bv\right) \nn \\
&+& \rho^{-1}\bB^*\cdot\left[F_{\bv}\times\left( \nabla\times G_{\bB^*}\right)-G_{\bv}\times \left(\nabla\times F_{\bB^*}\right)\right] \nn \\
&-& d_i \rho^{-1}\bB^*\cdot\left[\left(\nabla\times F_{\bB^*}\right)\times \left(\nabla\times G_{\bB^*}\right)\right]\nn \\
&+& d_e^2\rho^{-1}\left(\nabla\times \bv\right)\cdot \left[\left(\nabla\times F_{\bB^*}\right)\times \left(\nabla\times G_{\bB^*}\right)\right]\big\}\,, \nn
\end{eqnarray}
where $F_u:={\delta F}/{\delta u}$ denotes the functional derivative of $F$ with respect to the dynamical variable $u$.  The bracket of \eqref{poisson} generalizes the original MHD bracket of  \cite{morr_gree} by replacing $\bB$ by $\bB^*$ and the addition of the terms involving $d_e$ and $d_i$.

It is evident that the Poisson bracket above is antisymmetric and in \cite{abde_kawa_yosh} the authors proved by a tedious calculation (simplified in \cite{ling_morr_milo}) that it satisfies   the Jacobi identity. In view of \eqref{hamiltonian} and \eqref{poisson} the equations of motion can be cast into the following Hamiltonian form
\begin{equation}
\partial_t \mathbf{u}=\left\{\mathbf{u},\Hc\right\}\,,
 \label{ham_eq_mot_1}
\end{equation}
with $\mathbf{u}=\left(\rho,\bv,\bB^*\right)$. Equation \eqref{ham_eq_mot_1} can be written also as 
\begin{equation}
\partial_t u^i=J^{ij}\frac{\delta \Hc}{\delta u^j} \label{ham_eq_mot_2}\,,
\end{equation}
where $J^{ij}$ is the Poisson operator that defines the Poisson bracket according to  $\{F,G\}:=\langle  F_{u^i}, J^{ij}\,    G_{u^j}  \rangle$ with $\langle , \rangle$ being a pairing defined on the  phase(function)-space.
 A characteristic feature of Poisson brackets of the form \eqref{poisson} is that they have nontrivial kernels, i.e. there exist functionals $\Cc$ that  satisfy 
\begin{equation}
\left\{F,\Cc\right\}=0\, , \qquad \forall F \label{casimir_def}\,.
\end{equation}
Such functionals $\Cc$, called Casimirs, are global invariants of the dynamical evolution.  From \eqref{casimir_def} it is evident that the equations of motion are  unaffected by the addition of the Casimir invariants to the Hamiltonian $\Hc$. Therefore if we define a family of Hamiltonians,  $\mathfrak{F}=\Hc-\sum_i \Cc_i$ one can freely write 
\begin{equation}
\partial_t\mathbf{u}=\{\mathbf{u},\mathfrak{F}\}\,,
\end{equation}
instead of \eqref{ham_eq_mot_1} without changing the resultant equations of motion.   It is clear that stationary states are solutions of the equations $\{\mathbf{u},\mathfrak{F}\}=0$. Hence from Eq.~\eqref{ham_eq_mot_2} we understand that $\partial_t\mathbf{u}=0$ follows from  the vanishing of the first variation of the generalized Hamiltonian functional $\mathfrak{F}$ (the Energy-Casimir functional),  i.e., equilibrium states satisfy the condition
\begin{equation}
\delta \mathfrak{F}=
\delta\left(\Hc-\sum_i \Cc_i\right)=0 \,.
\label{en_cas}
\end{equation}
As pointed out in \cite{morr_rev,yos_morr}, in general not all equilibria emerge from such variations because of singularities in the Poisson bracket operator.

Regarding the dynamical evolution, the Casimirs play a topological role in the structure of the phase space, since the motion takes place on phase-space surfaces that are the Casimir level sets, commonly called symplectic leaves. Assigning initial values to the Casimirs is equivalent to the choice of a symplectic leaf,  i.e.,   a particular sub-space of the phase space on which the motion is restricted. The intersection with the energy level sets confines the trajectory of the dynamical evolution, and $\mathfrak{F}$ extremal points correspond to equilibrium states. We also note that the stability of  equilibrium points depends on the behavior of the second variation of the Energy-Casimir functional. 

The Casimir invariants for 3D barotropic XMHD have been calculated in \cite{abde_kawa_yosh} and \cite{ling_morr_milo}. The total mass, as it is expected, is conserved and also there exist two generalized Helicity-like invariants:
\begin{eqnarray}
\Cc_1&=&\int_Vd^3x\, \rho \,,\\
\Cc_{2,3}&=&\int_V d^3x\, \left(\bA^*+\lambda_{\pm}^{-1}\bv\right)\cdot \left(\bB^*+\lambda_{\pm}^{-1}\nabla\times\bv\right), \label{casimirs_xmhd}
\end{eqnarray}
with $\bB^*=\nabla\times\bA^*$ and $\lambda_{\pm}$ being the two roots of the quadratic equation $1-d_i\lambda-d_e^2\lambda^2=0$, that is $\lambda_\pm=({-d_i\pm\sqrt{d_i^2+4d_e^2}})/({2d_e^2})$. The two invariants in \eqref{casimirs_xmhd} have
forms similar to the canonical self-helicities of the twofluid model, which are composites of the fluid and magnetic momenta. However, they are not identical to the canonical helicities of the 2-fluid model since in the framework of XMHD, quasineutrality and smallness of the electron to ion mass ratio are assumed. The XMHD ``canonical momenta'' are proportional to $\bv+\lambda_\pm\bA$. Regarding the physical interpretation of  these invariants, in comparison to the  ordinary MHD magnetic helicity, a measure of the twist and linkage of magnetic flux tubes, these generalized helicities can be seen to measure the twist and linkage of the flux tubes of the generalized fields $\bB+\lambda^{-1}_{\pm}\nb\times\bv$ (or generalized vorticities). The two parameters $\lambda_\pm$ account for the differential motion of electrons and ions  departing  from  magnetic field lines.
 
\subsection{3D Energy-Casimir variational principle}
\label{3DECVP}

The Energy-Casimir variational principle, employing the general 3D expressions for the energy and the Casimir invariants, leads to equilibrium conditions   satisfied by the  magnetic and velocity fields. In the framework of single-fluid MHD the magnetic fields that  are  solutions of \eqref{en_cas} satisfy the so-called Beltrami condition: $\nabla\times \bB=\kappa \bB$, 
with the fluid velocity being parallel to $\bB$. In the context of XMHD, due to the form of the Hamiltonian and Casimir functionals, the magnetic and the velocity fields satisfy more complicated, coupled conditions that  allow more complex field configurations. These conditions can be derived from the vanishing of the  first variation of the Energy-Casimir functional $\mathfrak{F}$, i.e.,  by requiring  the vanishing of the coefficients of the arbitrary variations of $\rho$, $\bv$, and $\bB^*$,  
\begin{eqnarray}
&&\delta \mathfrak{F}=\delta\int_Vd^3x\,\Big\{\rho \frac{v^2}{2}+\rho U(\rho)+\frac{\bB\cdot\bB^*}{2}
-\alpha \rho 
\nn \\
&& -\beta_+ \left(\bA^*+\lambda_+^{-1}\bv\right)\cdot\left(\bB^*+\lambda_+^{-1}\nabla\times \bv\right)
 \nn \\
&&-\beta_-\left(\bA^*+\lambda_-^{-1}\bv\right)\cdot\left(\bB^*+\lambda_-^{-1}\nabla\times \bv\right)\Big\}=0\,,
 \label{3d_War_pri}
\end{eqnarray}
with  $\bB=\nabla\times\bA$ as usual and $\bA^*=\bA+d_e^3{\nabla\times \bB}/{\rho}$. The parameters $\alpha$ and  $\beta_\pm$  are Lagrangian multipliers with values  related to the total mass and   total  generalized helicities. Equation \eqref{3d_War_pri} leads to the following conditions
\begin{eqnarray}
&&\nabla\times \bB=2(\beta_++\beta_-)\bB^*
 \label{3d_magn_field}\\
&& \hspace{0.5cm} + 2\left(\beta_+ \lambda_+^{-1}+\beta_- \lambda_-^{-1}\right)\nabla\times \bv\, ,
\nn \\
&&\rho \bv= 2\left(\beta_+ \lambda_+^{-1}+\beta_- \lambda_-^{-1}\right)\bB^*
 \label{3d_Wel_field}\\
&&\hspace{0.5cm}  +2\left(\beta_+ \lambda_+^{-2}+\beta_-\lambda_-^{-2}\right)\nabla\times \bv \,,\nn\\
&&\frac{v^2}{2}+h(\rho)-d_e^2\, \frac{|\bJ|^2}{2\rho^2}+2\frac{d_e^2(\beta_+ +\beta_-)}{\rho^2}\bB^*\cdot \bJ
\nn \\
&& \hspace{0.5cm}  +2\frac{d_e^2(\beta_+\lambda_+^{-1}+\beta_-\lambda_-^{-1})}{\rho^2}\,\bJ\cdot(\nb\times \bv)
 =\alpha\,,
 \label{3d_Bernoulli}
\end{eqnarray}
where the enthalpy $h(\rho)=\left[\rho U(\rho)\right]_\rho$. The enthalpy  is related to pressure $P(\rho)$ through the following relation: 
\begin{equation}
h(\rho)=\int\frac{dP(\rho)}{\rho}\,.
\label{enthalpy}
\end{equation}
That the enthalpy $h$ of \eqref{enthalpy}  only depends on a single thermodynamics variable, the specifict volume $\rho^{-1}$, follows from the  barotropic assumption embodied in the choice of internal energy per unit mass $U$.  
For simplicity, our  Hamiltonian formulation was  restricted in this way,  but  it  can be generalized to include more  thermodynamic variables such as an entropy per unit mass (see, e.g., Ref.~\cite{amp2b} for MHD and Ref.~\cite{all} for XMHD).   A common  choice for this barotropic thermodynamic closure is the polytropic equation of state where $P(\rho)=\kappa\rho^{\gamma}$ with $\kappa$  constant (independent of  entropy).  With this choice 
\begin{equation}
h(\rho)=\frac{\gamma}{\gamma-1}\frac{P(\rho)}{\rho}\,.
\end{equation}
 From \eqref{3d_magn_field} and \eqref{3d_Bernoulli} we obtain the XMHD analogue of the Bernoulli equation, which reveals the distribution of the pressure, for velocity and magnetic field described by the mutual solutions of the coupled equations \eqref{3d_magn_field} and  \eqref{3d_Wel_field}:
\begin{equation}
\tilde{P}(\rho)=\alpha\rho-\rho\frac{v^2}{2}-\frac{d_e^2}{2\rho}|\nb\times \bB|^2\,,
\end{equation}
where $\tilde{P}={\gamma}P/({\gamma-1})$.

\section{Symmetric formulation via chain rule on functional derivatives}
\label{sec_III}

\subsection{Translationally symmetric Poisson bracket}

Assuming  continuous translational symmetry and adopting a Cartesian system $(x,y,z)$ the fields $\bB^*$ and $\bv$ can be written as follows: 
\begin{eqnarray}
\bB^*&=&B_z^*(x,y,t)\hat{z}+\nabla\psi^*(x,y,t)\times\hat{z}\,, \label{tran_sym_magn_field}\\
\bv&=&v_z(x,y,t)\hat{z}+\nabla\chi(x,y,t) \times \hat{z}+\nabla \Upsilon(x,y,t) \, ,
\label{tran_sym_vel_field}
\end{eqnarray}
where  $z$ is   the ignorable coordinate with corresponding unit vector $\hat{z}$  along the direction of translational invariance, $\psi^*$ is the poloidal flux function of $\bB^*$ and $\chi$, $\Upsilon$ are Clebsch-like potentials for  the poloidal velocity. The form of  Eq.~(\ref{tran_sym_magn_field}) ensures that the condition $\nabla\cdot\bB=0$ holds, while the existence of the term $\nabla\Upsilon$ in Eq.~(\ref{tran_sym_vel_field}) allows for  compressibility of the flow (provided  the function $\Upsilon(x,y)$ is not harmonic).  Upon setting $\Upsilon=0$ or $\Delta \Upsilon=0$ ($\Delta\equiv\nabla^2$,  the Laplacian), we can impose  incompressibility of the flow. Note that in view of the translational symmetry the representation  adopted for the velocity field is consistent with the Helmholtz decomposition theorem and hence it is generic for the description of any kind of symmetric flow. Taking the divergence and the curl of  Eqs.~(\ref{tran_sym_magn_field}) and  \eqref{tran_sym_vel_field} gives 
\begin{eqnarray}
\nabla\cdot\bv&=&\Delta\Upsilon\,, \\
\nabla\times \bv&=&-\Delta\chi \hz+\nb v_z \times \hz \,,\\
\nb\cdot \bB^*&=&0\,, \\
\nb\times \bB^*&=&-\Delta \psi^* \hz +\nb B_z^*\times \hz\,.
\end{eqnarray}
For convenience we define the following quantities:  $w:=\Delta \Upsilon$ or  $\Upsilon=\Delta^{-1}w$ and $\Omega=-\Delta \chi$  or $\chi=-\Delta^{-1}\Omega$. 

The transition from the general 3D Hamiltonian model to a translationally symmetric one  is accomplished by expressing the Hamiltonian \eqref{hamiltonian} and the Poisson bracket \eqref{poisson}, which are expressed in terms of the state vector $\mathbf{u}=\{\rho,\bv,\bB^*\}$, to those in terms of the symmetric state vector $\mathbf{u}_{TS}=\{\rho,v_z,\chi,\Upsilon,B_z^*,\psi^*\}$. This reduction of  phase space is achieved by   mapping  the functional derivatives with respect to the original variables  $\mathbf{u}$ to functional derivatives with respect to the variables  $\mathbf{u}_{TS}$.  This mapping is  computed using  the  chain rule for functional derivatives, obtained by equating first variations in terms of  the two sets of variables.  The variation of a functional $F[\rho,\bv,\bB^*]$ is 
\begin{equation}
\delta F[\mathbf{u}]=\int_Vd^3x\, \left(F_{\rho}\delta \rho+F_{\bv}\delta \bv+F_{\bB^*}\delta \bB^*\right)\,,
\end{equation}
while that in terms of  $\mathbf{u}_{TS}$  is 
\begin{eqnarray}
\delta F[\mathbf{u}_{TS}]&=&\int_D d^2x\, \Big[
F_{\rho}\delta\rho 
 + F_{v_z}\delta v_z+F_{\chi}\delta\chi +F_{\Upsilon}\delta\Upsilon
\nn\\
&&\hspace{1cm}+F_{B_z^*}\delta B_z^*+F_{\psi^*}\delta\psi^*
\Big]\,,
\label{FuTS}
\end{eqnarray}
where $D\subseteq R^2$ is a restriction of $V$ to $R^2$.  Using  $\delta\chi=-\Delta^{-1}\delta\Omega=-\Delta^{-1}\left(\hz\cdot\nb\times\delta\bv\right)$, $\delta\Upsilon=\Delta^{-1}\delta w=\Delta^{-1}\left(\nb\cdot \delta\bv\right)$,  and $\delta\psi^*=-\Delta^{-1}\left(\hz\cdot\nb\times\delta\bB^*\right)$,   \eqref{FuTS} can be rewritten as  
\begin{eqnarray}
\delta F&=& \int_Dd^2x\,\Big\{
 F_\rho \delta \rho +  F_{v_z}\hz\cdot \delta\bv-F_{\chi}\Delta^{-1}\left(\hz\cdot\nb\times\delta\bv\right)
\nn\\
&& \hspace{1cm}   +F_{\Upsilon}\Delta^{-1} \nb\cdot \delta\bv +F_{B_z^*}\hz\cdot\delta\bB^*
 \nn \\
&& \hspace{2cm}   -F_{\psi^*}\Delta^{-1}\left(\hz\cdot\nb\times\delta\bB^*\right)\Big\}\,.
\end{eqnarray}
Then, from  the self-adjointness of the operator $\Delta^{-1}$ and for appropriate boundary conditions such that the boundary terms vanish, we obtain 
\begin{eqnarray}
\delta F&=&\int_Dd^2 x \left(F_{\rho}\delta\rho+F_\bv\cdot \delta\bv+F_{\bB^*}\cdot \delta \bB^*\right)
\label{func_wer_relA}\\
&=& \int_Dd^2x\,\big[F_{\rho}\delta\rho+\left(F_{v_z}\hz+\nb F_{\Omega}\times \hz-\nb F_w\right)\cdot\delta\bv 
\nn\\
&&\hspace{1cm}+\left(F_{B_z^*} \hz-\nb(\Delta^{-1}F_{\psi^*})\times\hz\right)\cdot\delta\bB^*\big] \,,
\label{func_wer_rel}
\end{eqnarray}
where we have used the following relations:
\begin{equation}
F_\Upsilon=\Delta F_w\, , \qquad F_\chi=-\Delta F_\Omega\, ,
\end{equation}
which come from
\begin{eqnarray}
\int_D\!d^2x \, F_\Upsilon \delta\Upsilon&=&\!\int_D\!d^2x \, F_w\delta w=\int_D\!d^2x \, \Delta F_w \delta\Upsilon \,,\\
\int_D\!d^2x \, F_\chi \delta\chi&=&\!\int_D\!d^2x \, F_\Omega\delta \Omega=-\!\int_D\!d^2x \, \Delta F_\Omega \delta\chi\, ,
\end{eqnarray}
since the variations $\delta\chi$ and $\delta \Upsilon$ are arbitrary.
Upon comparing  \eqref{func_wer_relA} with  \eqref{func_wer_rel}, the following relations are deduced:
\begin{eqnarray}
F_{\bv}&=&F_{v_z}\hz+\nb F_\Omega \times\hz-\nb F_w \,,\label{f_w} \\
F_{\bB^*}&=&F_{B_z^*} \hz-\nb \left(\Delta^{-1}F_{\psi^*}\right)\times \hz\,, \label{f_B}\\
\nb\times F_{\bB^*}&=&F_{\psi^*}\hz+\nb F_{B_z^*}\times \hz \label{curl_f_B}\,.
\end{eqnarray}
Substituting Eqs.~\eqref{tran_sym_magn_field}, \eqref{tran_sym_vel_field}, \eqref{f_w}, \eqref{f_B},  and \eqref{curl_f_B} into the Poisson bracket of XMHD given by \eqref{poisson}, we obtain the translationally symmetric Poisson bracket of barotropic XMHD (see Appendix A for details):
\begin{eqnarray}
\{F,G\}^{^{XMHD}}_{_{TS}}&=&\int_Dd^2x\, \Big\{F_\rho\Delta G_w-G_\rho\Delta F_w  \nn \\
&&\hspace{-1.5cm} 
+\rho^{-1}\Omega\big(
\left[F_{\Omega},G_{\Omega}\right]+[F_w,G_w]
\nn\\
&& +\nb F_w \cdot\nb G_\Omega- \nb F_\Omega\cdot \nb G_w
\big) \nn \\
&&\hspace{-1.5cm} 
+v_z\big(
[F_\Omega,\rho^{-1} G_{v_z}]-[G_\Omega,\rho^{-1}F_{v_z}]
\nn\\
&& +\nb (\rho^{-1}G_{v_z}) \cdot \nb F_w-\nb (\rho^{-1}F_{v_z})\cdot \nb G_w  
 \nn \\ 
&& 
+   \rho^{-1}F_\Upsilon G_{v_z}-\rho^{-1}G_\Upsilon F_{v_z}
\big) \nn \\
&&\hspace{-1.5cm}
+\psi^*\big(
[F_{\Omega},\rho^{-1}G_{\psi^*}]-[G_{\Omega},\rho^{-1} F_{\psi^*}]
\nn\\
&&+[F_{B_z^*},\rho^{-1}G_{v_z}]-[G_{B_z^*},\rho^{-1}F_{v_z}]
  \nn \\
&& 
+ \nb F_w\cdot \nb(\rho^{-1}G_{\psi^*})-\nb G_w\cdot\nb (\rho^{-1}F_{\psi^*})
\nn\\
&& +\rho^{-1}F_\Upsilon G_{\psi^*}-\rho^{-1}G_\Upsilon F_{\psi^*}
\big) \nn \\
&&\hspace{-1.5cm}
+\rho^{-1}B_z^*\big(
 [F_\Omega ,G_{B_z^*}]-[G_\Omega , F_{B_z^*}] 
\nn\\
&&+\nb F_w\cdot\nb G_{B_z^*}  -\nb G_w\cdot\nb F_{B_z^*} 
\big) \nn \\
&&\hspace{-1.5cm}
+d_i \psi^* \big([G_{B_z^*},\rho^{-1}F_{\psi^*}]-[F_{B_z^*},\rho^{-1}G_{\psi^*}]
\big) 
\nn\\
&&\hspace{-1.5cm} -d_i \rho^{-1} B_z^*[F_{B_z^*},G_{B_z^*}]
+d_e^2 \rho^{-1}\Omega [F_{B_z^*},G_{B_z^*}]
\nn\\
&&\hspace{-1.5cm} +d_e^2v_z\big([F_{B_z^*},\rho^{-1}G_{\psi^*}]-[G_{B_z^*},\rho^{-1}F_{\psi^*}]\big)
\Big\} \,,
 \label{ts_poisson}
\end{eqnarray}
where  $[a,b]:=\left(\nb a\times\nb b\right)\cdot\hz=(\partial_x a)(\partial_y b)-(\partial_x b) (\partial_y a)$.  Here  we exploited the identity 
\begin{equation}
\int_Dd^2x\, [a,b]c=\int_Dd^2x\, [c,a]b=\int_Dd^2x\,[b,c]a\,,
 \label{identity}
\end{equation}
which holds for arbitrary functionals $a,b,c$ under appropriate boundary conditions (e.g.\  periodic boundary conditions). 

The antisymmetry of the bracket \eqref{ts_poisson} follows naturally from the antisymmetry of the  Poisson bracket  $[a,b]=-[b,a]$ and the vanishing of the boundary terms arising  from integration by parts and Gauss' theorem.  The Jacobi identity of \eqref{ts_poisson}  follows because of the reduction procedure. 
Similarly, by  substitution  the  symmetric representation of the Hamiltonian is given by
\begin{eqnarray}
\Hc_{_{TS}}^{^{XMHD}}&=&\frac{1}{2}\int_Dd^2x\, \bigg\{\rho\left(v_z^2+|\nabla \chi|^2+|\nb\Upsilon|^2\right)
\label{ts_hamiltonian}\\
&&+2\rho \left([\Upsilon,\chi]+U(\rho)\right) +B_z^2+|\nb\psi|^2 \nn \\
&&+\frac{d_e^2}{\rho}\left[\left(\Delta\psi\right)^2+|\nb B_z|^2\right]\bigg\}\nn \\
&=&\int_Dd^2x\, \bigg\{\frac{\rho}{2}\left(v_z^2+|\nabla \chi|^2+|\nb\Upsilon|^2\right)
\nn\\
&&+\rho \left([\Upsilon,\chi]+U(\rho)\right)
+\frac{B_z^*B_z}{2}+\frac{\nb\psi^*\cdot\nb\psi}{2} \bigg\}\,.
\nn
\end{eqnarray}
With \eqref{ts_hamiltonian} the  translationally symmetric equations of motion take the form   
$\partial_t \mathbf{u}_{TS}=\{\mathbf{u}_{TS},\Hc_{TS}\}^{^{XMHD}}_{_{TS}}$. The bracket \eqref{ts_poisson} has a more complicated form than its MHD counterpart obtained in \cite{andr_morr_pego}, due to the terms that  originate from the ion and electron contributions, having coefficients $d_i$ and $d_e$, respectively. However, a  remarkable transformation  introduced in \cite{ling_morr_milo} can simplify it. The new transformed bracket has the form of the translationally symmetric HMHD Poisson bracket, which can be obtained by setting $d_e=0$ in \eqref{ts_poisson}, but with dependence  on a generalized magnetic field variable 
\begin{equation}
\bB_\pm=\bB^*+\lambda_{\pm}^{-1}\nabla\times \bv \,.
\end{equation} 
The new magnetic field variable $\bB_\pm$, in view of Eqs.~\eqref{tran_sym_magn_field} and \eqref{tran_sym_vel_field} can be written as
\begin{eqnarray}
\bB_\pm&=&\bB^*+\lambda_\pm^{-1}\nb\times\bv
\nn\\
&=&\left(B_z^*+\lambda_{\pm}^{-1} \Omega\right)\hz+\nb\left(\psi^*+\lambda_{\pm}^{-1}v_z\right)\times \hz
\nn\\
&=&B_{z}^\pm\hz+\nb\psi_\pm\times\hz\,,
\end{eqnarray}
i.e. we have 
\begin{equation}
B_z^\pm=B_z^*+\lambda_{\pm}^{-1} \Omega, \qquad \psi_\pm=\psi^*+\lambda_{\pm}^{-1}v_z \label{rem_trans_1}\,.
\end{equation}
We can prove that under the change
\[
\{\rho,v_z,\Omega,\Upsilon,B_z^*,\psi^*\}\leftrightarrow \{\rho,v_z,\Omega,\Upsilon,B_z^\pm,\psi_\pm\}
\]
 the functional derivatives change as follows:
\begin{eqnarray}
&& F_{v_z}\rightarrow F_{v_z}+\lambda^{-1}_{\pm}F_{\psi_\pm}, \quad F_\Omega\rightarrow F_\Omega+\lambda^{-1}_\pm F_{B_z^\pm},   \label{rem_trans_2}\\
&&F_\Upsilon\rightarrow F_\Upsilon,\quad  F_{\psi^*}\rightarrow F_{\psi_\pm}, \quad F_{B_z^*}\rightarrow F_{B_z^\pm}\,,
\nn
\end{eqnarray}
with the change of  variables of  \eqref{rem_trans_1}.  Upon inserting the transformation of  the functional derivatives  of  \eqref{rem_trans_2} into \eqref{ts_poisson} we obtain the following bracket:
\begin{eqnarray}
\{F,G\}^{^{XMHD}}_{_{TS}}&=&\int_Dd^2x\, \Big\{ F_\rho\Delta G_w-G_\rho\Delta F_w 
 \label{ts_poisson_sim} \\
&&\hspace{-1.25cm} +\rho^{-1}\Omega\Big(
\left[F_{\Omega},G_{\Omega}\right] +[F_w,G_w] 
\nn\\
&&\hspace{-.5cm}+\nb F_w \cdot\nb G_\Omega- \nb F_\Omega\cdot \nb G_w
\Big) \nn \\
&&\hspace{-1.25cm} +v_z\Big(
[F_\Omega,\rho^{-1} G_{v_z}]-[G_\Omega,\rho^{-1}F_{v_z}] \nn \\
&&\hspace{-.5cm}+ \nb (\rho^{-1}G_{v_z})\cdot \nb F_w-\nb (\rho^{-1}F_{v_z})\cdot \nb G_w 
\nn\\
&& \hspace{-.5cm}+  \rho^{-1}F_\Upsilon G_{v_z}-\rho^{-1}G_\Upsilon F_{v_z}
\Big) \nn \\
&&\hspace{-1.25cm} +\psi_\pm\Big(
[F_{\Omega},\rho^{-1}G_{\psi_\pm}]-[G_{\Omega},\rho^{-1} F_{\psi_\pm}] 
\nn\\
&&\hspace{-.5cm}+[F_{B_z^\pm},\rho^{-1}G_{v_z}]-[G_{B_z^\pm},\rho^{-1}F_{v_z}] \nn \\
&&  \hspace{-.5cm}+ \nb F_w\cdot \nb (\rho^{-1}G_{\psi_\pm}) - \nb G_w\cdot \nb (\rho^{-1}F_{\psi_\pm})
\nn\\
&&\hspace{-.5cm}+\rho^{-1}F_\Upsilon G_{\psi_\pm}-\rho^{-1}G_\Upsilon F_{\psi_\pm}
\Big) \nn \\
&&\hspace{-1.25cm} +\rho^{-1}B_z^\pm\Big([F_\Omega ,G_{B_z^\pm}] - [G_\Omega , F_{B_z^\pm}] 
\nn\\
&&\hspace{-.5cm}+\nb F_w\cdot\nb G_{B_z^\pm}-\nb G_w\cdot\nb F_{B_z^\pm}
 \Big) \nn \\
&&\hspace{-1.25cm} -\nu_{\pm} \rho^{-1} B_z^\pm[F_{B_z^\pm},G_{B_z^\pm}]
\nn\\
&&\hspace{-1.25cm}+ \nu_{\pm} \psi_\pm \left([G_{B_z^\pm},\rho^{-1}F_{\psi_\pm}]-[F_{B_z^\pm},\rho^{-1}G_{\psi_\pm}]\right) 
\Big\} \, ,
\nn
\end{eqnarray}
where $\nu_{\pm}:=d_i-2\lambda_\pm^{-1}$.

As was the case for \eqref{ts_poisson},  the bracket \eqref{ts_poisson_sim} with the Hamiltonian \eqref{ts_hamiltonian}, generate the  translationally symmetric XMHD equations of  motion according to $\partial_t \mathbf{u}_{TS}=\{\mathbf{u}_{TS},\Hc_{TS}\}^{^{XMHD}}_{_{TS}}$.

\subsection{Translationally symmetric Casimirs}
\label{ssec:transCas}

As in the 3D case, there exist Casimir invariants conserved by the translationally symmetric dynamics.   As already mentioned, the Casimirs satisfy $\{F,\Cc\}=0$, $\forall F$.  For the bracket \eqref{ts_poisson_sim} this gives 
\begin{eqnarray}
&&\int_Dd^2x\,\big(F_\rho \Qc_1+F_{v_z}\Qc_2+F_{\Omega}\Qc_3+F_w\Qc_4
\nn\\
&& \hspace{1.5cm} +F_{B_z^\pm}\Qc_5+F_{\psi_\pm}\Qc_6
\big)=0\, ,
 \label{cas_wet_in}
\end{eqnarray}
where the quantities $\Qc_i$ $(i=1,2,...,6)$ are given by the following expressions: 
\begin{eqnarray}
\Qc_1&=&\Delta \Cc_w=\Cc_\Upsilon \,,\label{cas_wet_1}\\
\Qc_2&=&[\Cc_\Omega,v_z]+\nabla \cdot (v_z\nb \Cc_w)-v_z \Cc_\Upsilon-[\psi_\pm,\Cc_{B_z^\pm}]\,, \label{cas_wet_2}\\
\Qc_3&=&\nb\cdot (\rho^{-1}\Omega\nb\Cc_w)- [\rho^{-1}\Omega,\Cc_{\Omega}]\label{cas_wet_3}\\
&& -[v_z,\rho^{-1}\Cc_{v_z}] -[\psi_\pm,\rho^{-1}\Cc_{\psi_\pm}]-[\rho^{-1}B_z^\pm,\Cc_{B_z^\pm}] \,,
\nn \\
\Qc_4&=&\Delta \left(\rho^{-1}v_z \Cc_{v_z}\right)+ \Delta\left(\rho^{-1}\psi_\pm\Cc_{\psi_\pm}\right) -\Delta \Cc_\rho
\nn\\
&&  -[\rho^{-1}\Omega,\Cc_w]   -\nabla \cdot \big(\rho^{-1}\Omega\nb \Cc_\Omega +v_z\nb(\rho^{-1}\Cc_{v_z})
 \nn\\
 &&\hspace{.5cm} +\psi_\pm\nb(\rho^{-1}\Cc_{\psi_\pm})+ \rho^{-1}B_z^\pm\nb \Cc_{B_z^\pm}
 \big)\,,
 \label{cas_wet_4} \\ 
\Qc_5&=&[\rho^{-1}\Cc_{v_z},\psi_\pm]+[\Cc_\Omega,\rho^{-1}B_z^\pm] +\nb\cdot (\rho^{-1}B_z^\pm\nb \Cc_w)
\nn\\
&&+\nu_{\pm} [\psi_\pm,\rho^{-1}\Cc_{\psi_\pm}]+\nu_{\pm} [\rho^{-1}B_z^\pm,\Cc_{B_z^\pm}]\,,\label{cas_wet_5}\\
\Qc_6&=& [\Cc_\Omega,\psi_\pm]+\nb \cdot (\psi_\pm \nb \Cc_w)-\psi_\pm \Cc_{\Upsilon}
\nn\\
&&+\nu_{\pm} [\psi_\pm,\Cc_{B_z^\pm}] \, .\label{cas_wet_6}
\end{eqnarray}
For \eqref{cas_wet_in} to be satisfied for arbitrary variations, the coefficients $\Qc_i$ must  vanish separately, i.e., 
\begin{equation}
\Qc_i=0\,, \qquad  i=1,2,...,6\, . \label{cas_wet_sys}
\end{equation}
Equation $\Qc_1=0$, i.e. $\Cc_\Upsilon=0$, implies that the Casimirs are independent of $\Upsilon$. Equations $\Qc_4=0$ and $\Qc_3=0$ are,  respectively,  the divergence and the z component of the curl of the following equation:
\begin{eqnarray}
&&\nb \left(\rho^{-1}v_z \Cc_{v_z}\right)+\nb\left(\rho^{-1}\psi_\pm\Cc_{\psi_\pm}\right)-\nb \Cc_\rho
\label{curl_wiv}\\
&&\hspace{.5cm}  -v_z \nb (\rho^{-1}\Cc_{v_z})-\psi_\pm
\nb (\rho^{-1} \Cc_{\psi_\pm})-\rho^{-1}B_z^\pm \nb \Cc_{B_z^\pm}
\nn \\
&&\hspace{1cm}-\rho^{-1}\Omega \nb \Cc_w\times \hz-\rho^{-1} \Omega \nb \Cc_\Omega
=0\,. \nn
\end{eqnarray}
We observe that  \eqref{curl_wiv} is satisfied automatically for $\Cc_\rho$=const.,  which gives the first Casimir, 
\begin{equation}
\Cc_m=\int_Dd^2x\, \rho \,.
 \label{mass_cas}
\end{equation}
Note that in general a solution to $\Qc_4=0$ could be satisfied by $\Cc_m=\int_Dd^2x\, \rho\Phi$ with $\Phi$ being a harmonic function,   $\Delta\Phi=0$. 
The equations $\Qc_2=0$ and  $\Qc_6=0$ can be combined by multiplying the first by $\nu_{\pm}$ and adding it to the second, 
\begin{eqnarray}
\nu_{\pm} \Qc_2+\Qc_6&=& [\Cc_\Omega,\psi_\pm +\nu_{\pm} v_z]
\label{eq_for_cas_1}\\
&&\hspace{.5cm} +\nb\cdot [(\psi_\pm+\nu_{\pm} v_z)\nb \Cc_w]=0 \,,
\nn
\end{eqnarray}
where we have used that $\Cc_\Upsilon=0$. With the  new variable  $\xi_\pm=\psi_\pm+\nu_{\pm}v_z$, 
 \eqref{eq_for_cas_1} becomes
\begin{equation}
[\xi_\pm,\Cc_\Omega]-\nb \cdot (\xi_\pm \nb \Cc_w)=0\,. 
\end{equation}
Equivalently,  we can write 
\begin{equation}
 \nb \Cc_\Omega-\hz\times \nb \Cc_w=\xi_\pm^{-1}\nb A_\pm \,,\label{grad_omeg}
\end{equation}
for $\xi_\pm\neq 0$ and $A_\pm$ being arbitrary functions. The $z$-component of the curl of Eq.~\eqref{grad_omeg} is 
\begin{equation}
\Delta \Cc_w=\Cc_\Upsilon=[A_\pm,\xi_\pm^{-1}]=0\,. 
\end{equation}
Therefore,  the  $A_\pm$ are arbitrary functions of  $\xi_\pm$, i.e.
\begin{equation}
A_\pm=A_\pm(\psi_\pm+\nu_\pm v_z)\,.
\end{equation}
Now   divergence of   \eqref{grad_omeg} translates to
\begin{equation}
\Delta \Cc_\Omega =\nb \cdot (\xi_\pm^{-1}\nb A_\pm)
=\nb \cdot (A_\pm'\xi^{-1}\nb \xi_\pm)=\Delta \Ac_\pm\,,
 \label{delta_Omega}
\end{equation}
where the functions $\Ac_\pm$ are related to $A_\pm$ via $\Ac_\pm':=\xi_\pm^{-1}A_\pm'$. 
According to  \eqref{delta_Omega}, we have $\Cc_\Omega= \Ac_\pm$, up to a harmonic function, therefore
\begin{eqnarray}
\Cc=\int_Dd^2x\,  \Omega \Ac_\pm(\psi_\pm+\nu_\pm v_z)+\Fc(B_z^\pm,\psi_\pm,v_z)\label{c_omega} \,.
\end{eqnarray}
Inserting \eqref{c_omega} into $\Qc_6=0$ we obtain 
\begin{eqnarray}
[\Ac_\pm,\psi_\pm {}]+\nu_{\pm}[\psi_\pm,\Fc_{B_z^\pm}]=0\,.
 \label{cas_A_3}
\end{eqnarray}
From  \eqref{cas_A_3}  we derive  $\Fc_{B_z^\pm}=\nu_\pm^{-1}\Ac_\pm+\Gc_\pm(\psi_\pm)$ , with $\Gc_\pm$ being an arbitrary function of $\psi_\pm$,  which combined with  \eqref{c_omega}, gives the following families of solutions
\begin{eqnarray}
\Cc_1^\pm&=&\int_Dd^2x\, \left(B_z^\pm+\nu_\pm \Omega\right)\Ac_\pm(\psi_\pm+\nu_\pm v_z)
\nn\\
&&\hspace{3 cm} +\tilde{\Fc}_\pm(v_z,\psi_\pm) \,,
\label{cas_O_B_F}
\\
\Cc_2^\pm&=&\int_Dd^2x\, B_z^\pm \Gc_\pm(\psi_\pm) +\tilde{\Fc}_\pm(v_z,\psi_\pm) \label{cas_B_G}\,,
\end{eqnarray}
for $\Gc_\pm = 0$ and $\Ac_\pm =0$ respectively. We remark here that if after Eq.~\eqref{delta_Omega} one take $\Cc_\Omega=\Ac_\pm+\Phi(x,y)$, with $\Phi(x,y)$ being a harmonic function, then it is not difficult to prove that the additional functional, coming from $\Phi$, will be a Casimir only if $\Phi=\Phi(\psi_\pm+\nu_\pm v_z)$ or $\Phi=\Phi(\psi_\pm)$. This would result in special cases of $\Cc_1^\pm$ and $\Cc_2^\pm$, which may be valid if the motion of the variables $\psi_\pm$ and $v_z$ is restricted by a differential constraint. Having found the dependencies of the Casimir invariants on $\Omega$ and $B_z^\pm$, it remains to investigate any additional dependencies on $v_z$ and $\psi_\pm$, represented by $\tilde{\Fc}$.  Upon substituting   \eqref{cas_O_B_F} into $\Qc_5=0$,  the latter reduces to 
\begin{equation}
[\rho^{-1} (\tilde{\Fc}_{v_z}-\nu_\pm \tilde{\Fc}_{\psi_\pm}),\psi_\pm]=0\,,
 \label{cas_vz_psi}
\end{equation}
which additionally gives the following functionals:
\begin{eqnarray}
\Cc_3^\pm&=&\int_Dd^2x\, \rho \Kc_\pm(\psi_\pm+\nu_\pm v_z)\,, \label{cas_K}\\
\Cc_4^\pm&=&\int_Dd^2x\, \rho \Mc_\pm(\psi_\pm)\label{cas_M}\,,
\end{eqnarray}
where $\Mc_\pm$ and $\Kc_\pm$ are arbitrary functions. The functionals above express the conservation of canonical-like momenta in the direction of symmetry. Also they encapsulate  conservation of mass, since $\Cc_m$ is the special case with $\Kc_\pm=1$, and the conservation of the mechanical momentum along the axis of symmetry. To make this clear,  note that $\Cc_4$ for example, if $\Mc_\pm$ is differentiable, can be written as  $\Cc_4=\int_Dd^2x\,\rho\int_1^{\psi_\pm}ds\,\Nc_\pm(s)$ with $\Nc_\pm(\psi_\pm)=\Mc_\pm'(\psi_\pm)$. Under a change of the integration variable this takes the form $\Cc_4=\int_Dd^2x\,\rho\int_1^{v_z}ds\Nc_\pm(\psi_\pm+\nu_\pm v_z-\nu_\pm s)$. For $\Nc_\pm=1$ we recover the conservation of mechanical momentum along the $z$-axis. Notice that in view of \eqref{cas_K} and \eqref{cas_M}, the term $\tilde{\Fc}_\pm$ in \eqref{cas_O_B_F} and \eqref{cas_B_G} can be subtracted. It is not difficult to verify that $\Cc_1^\pm, \, \Cc_2^\pm, \, \Cc_3^\pm , \, \Cc_4^\pm$ satisfy $\Qc_{2,3,4}=0$ as well and therefore all Casimir-determining equations \eqref{cas_wet_sys} are satisfied. Also, since $\psi_\pm=\psi_\mp+\nu_\mp v_z$ and $B_z^\pm=B_z^\mp+\nu_\mp\Omega$ (because $\lambda_\pm^{-1}=d_i-\lambda_\mp^{-1}$), the functionals \eqref{cas_O_B_F}--\eqref{cas_M}, represent just four independent families of invariants. Therefore one may freely keep either the set denoted by $(+)$ or the $(-)$ representation. In terms of the original magnetic variables $(B_z^*,\psi^*)$ the XMHD Casimir invariants are written as:
\begin{eqnarray}
\Cc_1&=&\int_Dd^2x\,(B^*_z+\mu\Omega)\Ac(\psi^*+\mu v_z) \,,\label{xcas_1-}\\
\Cc_2&=&\int_Dd^2x\,(B^*_z+\lambda^{-1}\Omega)\Gc(\psi^*+\lambda^{-1} v_z)\,, \label{xcas_2-}\\
\Cc_3&=&\int_Dd^2x\, \rho \Kc (\psi^*+\mu v_z)\,, \label{xcas_3-} \\
\Cc_4&=&\int_Dd^2x\, \rho \Mc(\psi^*+\lambda^{-1}v_z) \label{xcas_4-}\,,
\end{eqnarray} 
where the parameters $\lambda$ and $\mu$ are either $(\lambda,\mu)=(\lambda_+,\mu_+)$ or $(\lambda,\mu)=(\lambda_-,\mu_-)$, with $\mu_\pm:=\nu_\pm+\lambda_\pm^{-1}=d_i-\lambda_\pm^{-1}=\lambda_\mp^{-1}$.  

As discussed above, the Casimirs $\Cc_{3,4}$ express the conservation of mass and the conservation of (canonical) momenta in the direction of symmetry. In addition the Casimirs $\Cc_{1,2}$ are the symmetric counterparts of the generalized helicities \eqref{casimirs_xmhd}. Unlike the 3D Casimirs, the symmetric invariants form infinite families, due to the existence of the arbitrary functions $\Ac,\,,\Gc,\, \Kc,\, \Mc$. Later we will see that these arbitrary functions are transferred, by the variational principle, into the equilibrium equations giving in principle the possibility of  constructing  infinitely many classes of equilibria, unlike the 3D case where all equilibria obtained from an  energy-Casimir variational principle belong to the same class (see Eqs.~\eqref{3d_magn_field}-\eqref{3d_Bernoulli}).


\subsection{Hall MHD limit}
\label{ssec:HMHDlim}
Hall-MHD neglects electron inertia and therefore is recovered by the XMHD model for $d_e\rightarrow 0$.     If we assume $(\lambda,\mu)=(\lambda_+,\mu_+)$ and take the limit $d_e\rightarrow 0$,  then $B_z^*\rightarrow B_z$,  $\psi^*\rightarrow \psi$, $\lambda^{-1}\rightarrow d_i$ and $\mu\rightarrow 0$. In  this case the Hamiltonian becomes identical in form to the ordinary MHD symmetric Hamiltonian, that is
\begin{eqnarray}
\Hc_{_{TS}}^{^{HMHD}}&=&\frac{1}{2}\int_D d^2x\, \Big\{\rho\left(v_z^2+|\nabla \chi|^2+|\nb\Upsilon|^2\right)
\nn\\
&&\hspace{.5cm}+2\rho \left([\Upsilon,\chi]+U(\rho)\right) 
+B_z^2+|\nb\psi|^2 \Big\}\,.
\nn
\end{eqnarray} 
Also the HMHD Casimir invariants are
\begin{eqnarray}
\Cc_1^{^{HMHD}}&=&\int_Dd^2x\,B_z\Ac(\psi )\,,
 \label{hmhd_casimir1} \\
\Cc_2^{^{HMHD}}&=&\int_Dd^2x\,(B_z+d_i\Omega)\Gc(\psi+d_i v_z)\,,
 \label{hmhd_casimir2}\\
\Cc_3^{^{HMHD}}&=&\int_Dd^2x\, \rho \Kc (\psi)\,,
 \label{hmhd_casimir3}\\
\Cc_4^{^{HMHD}}&=&\int_Dd^2x\, \rho \Mc(\psi+d_i v_z) \,.
 \label{hmhd_casimir4}
\end{eqnarray}
If we return to the symmetric XMHD Poisson bracket and set $d_e=0$, we can verify that the HMHD bracket possess the Casimirs \eqref{hmhd_casimir1}--\eqref{hmhd_casimir4}. We remark that the ``generalized variables" $B_z+d_i\Omega$, $\psi+d_iv_z$ appear in \eqref{hmhd_casimir1}--\eqref{hmhd_casimir4} since the ion canonical helicity $\int_Vd^3x\, (\bA+d_i\bv)\cdot(\bB+d_i\nb\times\bv)$ is a Casimir invariant in 3D HMHD \cite{ling_morr_milo}.

\subsection{MHD limit}
\label{ssec:MHDlim}

For the MHD limit we additionally require $d_i\rightarrow 0$ in \eqref{hmhd_casimir1}-\eqref{hmhd_casimir4}, which yields only two  of the translationally symmetric ideal MHD  Casimir invariants of   \cite{andr_morr_pego,amp1,moawad}.   However, it was observed in the first Hamiltonian structure that contained Hall physics \cite{HHM},  that care must be taken with this limit  (see also \cite{gras_tass_abde_morr}), which appears at face value to not obviously yield the MHD versions of the Casimirs $\Cc_2^{^{HMHD}}$ and  $\Cc_3^{^{HMHD}}$. 
 
To see how this transpires, we  rewrite the invariants $\Cc_2^{^{HMHD}}$  and  $\Cc_4^{^{HMHD}}$ as follows:   
\begin{eqnarray}
\Cc_2^{^{HMHD}}&=&\int_Dd^2x \,d_i^{-1} \big(B_z+d_i\Omega\big)
\nn\\
&&\hspace{.5cm}  \times\big[\Gc(\psi) +d_i v_z \Gc'(\psi)+ \Oc(d_i^{2})\big]
\nn\\
&=&\int_Dd^2x\, \big[ d_i^{-1} B_z \Gc(\psi) + \Omega \Gc(\psi)  + v_z B_z \Gc'(\psi) 
\nn\\
&& \hspace{.5cm}  +\  d_i\Omega v_z \Gc'(\psi) + \Oc(d_i)\big]\,,
\label{xcas_1_2}
\\\
\Cc_4^{^{HMHD}}&=&\int_Dd^2x\, \rho \,d_i^{-1}\big[ \Mc(\psi) 
 +\  d_i v_z\Mc'(\psi) \nn\\
&& \hspace{.5cm}+ \Oc(d_i^{2})\big]
 \label{xcas_3-2}\,,
\end{eqnarray}
where we have scaled the arbitrary functions $\Gc$ and $\Mc$ by a factor of $d_i$.  
If we then take $d_i\rightarrow 0$ the first term of  $\Cc_2^{^{HMHD}}$ in (\ref{xcas_1_2}) is  seen  to  diverge.  However, this term  is itself a special case of $\Cc_1^{^{HMHD}}$, so it can be subtracted from \eqref{xcas_1_2}, giving
\begin{eqnarray}
\Cc_2^{^{MHD}}&=&\int_Dd^2x\, \big(  \Omega \Gc(\psi)  + v_z B_z \Gc'(\psi)  \big) 
\nn\\
&=&\int_Dd^2x\, \big(\nabla \chi \cdot \nabla\psi  + v_z B_z \big)  \Gc'(\psi) \,.
\label{xcas_1_lim}
\end{eqnarray}
A similar argument applies for the limit of the Casimir $\Cc_4$ of \eqref{xcas_3-2}. Therefore in the MHD limit $d_i\rightarrow 0$  all Casimirs approach their translationally symmetric MHD counterparts of \cite{andr_morr_pego,amp1}. To summarize, (with a redefinition of the arbitrary functions $\Gc$ and $\Mc$)   the following  translationally symmetric MHD Casimirs are obtained from the XMHD Casimirs in the limit $d_e\rightarrow 0$ followed by  $d_i\rightarrow 0$: 
\begin{eqnarray}
\Cc_{1}^{^{MHD}}&=&\int_D d^2x\, B_z{\Ac}(\psi) \,,
\label{MHDcas1}\\
\Cc_{2}^{^{MHD}}&=&\int_Dd^2x\, (B_zv_z+\nabla\psi\cdot\nabla\chi ){\Gc}(\psi)\,,
\label{MHDcas2}\\
\Cc_{3}^{^{MHD}}&=&\int_Dd^2x\, \rho  \Kc(\psi)\,,
\label{MHDcas3}\\
\Cc_4^{^{MHD}}&=&\int_D d^2x\,\rho v_z \Mc(\psi)\,.
\label{MHDcas4}
\end{eqnarray}

 Note that the Casimirs $\Cc_1^{^{MHD}}$, $\Cc_3^{^{MHD}}$ are identical  to the HMHD Casimir functionals given by \eqref{hmhd_casimir1}, \eqref{hmhd_casimir3}. This follows from the fact that the magnetic helicity is a common Casimir invariant for both models. The MHD limit of the HMHD model is also discussed in \cite{yosh_hame,hame_1}, although it is not shown how to limit the HMHD Casmirs  into their MHD values.

\subsection{Inertial MHD limit}

Inertial MHD (IMHD) occurs upon setting $d_i=0$ while $d_e\neq 0$, the reverse of the limit of Sec.~\ref{ssec:HMHDlim}.  IMHD is valid when the characteristic time scale for changes in the  current $\bJ$ is significantly shorter than the electron gyro-period \cite{kimu_morr}.  The Hamiltonian of translationally symmetric IMHD is $\Hc_{_{TS}}^{^{IMHD}}=\Hc_{_{TS}}^{^{XMHD}}$,  as  given by (\ref{ts_hamiltonian}).  
In the inertial MHD limit $d_i\rightarrow 0$ the parameters $\lambda_\pm=({-d_i\pm\sqrt{d_i^2+4d_e^2}})/({2d_e^2})$ go to $\pm d_e^{-1}$ and hence $\lim_{d_i\rightarrow 0}\mu_{\pm} = \mp d_e$,  which leads to the following form for the Casimir invariants:
\begin{eqnarray}
\Cc_1^{^{IMHD}}&=&\int_Dd^2x\,(B_z^*+d_e\Omega)\Ac(\psi^* +d_e v_z)\,,
 \label{imhd_casimirs1}\\
\Cc_2^{^{IMHD}}&=&\int_Dd^2x\,(B_z^*-d_e \Omega)\Gc(\psi^*-d_e v_z)\,,
 \label{imhd_casimirs2}\\
\Cc_3^{^{IMHD}}&=&\int_Dd^2x\, \rho \Kc (\psi^*+d_e v_z)\,,
 \label{imhd_casimirs3} \\
\Cc_4^{^{IMHD}}&=&\int_Dd^2x\, \rho \Mc(\psi^*-d_e v_z) \,.
 \label{imhd_casimirs4}
\end{eqnarray}
Upon taking $d_e\rightarrow 0$ in a manner similar to the $d_i$ limits of  Sec.~\ref{ssec:MHDlim},  one can show that the Casimirs of \eqref{imhd_casimirs1}--\eqref{imhd_casimirs4} become the MHD Casimirs of  \eqref{MHDcas1}--\eqref{MHDcas4}.  For example, upon setting $\Kc=\Mc$,  $\lim_{d_e\rightarrow 0}(\Cc_3^{^{IMHD}}\!\!\!-\,\, \Cc_4^{^{IMHD}})/d_e$ becomes  $\Cc_4^{^{MHD}}$\!\!.  The Casimir $\Cc_2^{^{MHD}}$ follows similarly.  

 An interesting property of IMHD is that the well-known MHD cross helicity  is also a Casimir for  IMHD if  $\bB\rightarrow\bB^*$ \cite{ling_morr_tass}, that is 
\begin{equation}
\Cc_c=\int_Vd^3x\,\bv\cdot\bB^*\,, \label{c_h_imhd}
\end{equation}
is a Casimir invariant of the general 3D IMHD model. For a translationally symmetric system, inserting the representations of  \eqref{tran_sym_magn_field} and  \eqref{tran_sym_vel_field} into  \eqref{c_h_imhd} and assuming appropriate boundary conditions, the symmetric version of the functional above is 
\begin{equation}
\Cc_c=\int_Dd^2x \, \left(v_zB_z^*+\Omega\psi^*\right) \, , \label{sym_c_h_imhd}
\end{equation}
which at a first glance is not included in \eqref{imhd_casimirs1}--\eqref{imhd_casimirs4}. However it is easy to see that upon choosing $\Ac=\psi^*+d_e v_z$ and $\Gc=\psi^*-d_ev_z$ , the Casimir \eqref{sym_c_h_imhd} is recovered from  $(\Cc_1^{^{IMHD}}-\Cc_2^{^{IMHD}})/(2d_e)$.
\section{Energy-Casimir variational principle with symmetry \label{sec_IV}}

\subsection{The variational principle}

Having determined the invariants of the translationally symmetric XMHD we can  construct easily 
 the Energy-Casimir variational principle of \eqref{en_cas} for XMHD equilibria that have  translation symmetry. Similar variational principles with symmetry can be found in \cite{andr_morr_pego,amp1,moawad,andr_pego}.  Gathering together 
 relations \eqref{ts_hamiltonian}  and \eqref{xcas_1-}--\eqref{xcas_4-} the Energy-Casimir principle $\delta \mathfrak{F}=0$ reads as follows:
 \begin{eqnarray}
&&\delta\int_Dd^2x\, \bigg\{\frac{\rho}{2}\left(v_z^2+|\nabla \chi|^2+|\nb\Upsilon|^2\right)+\rho \left([\Upsilon,\chi]+U(\rho)\right)
\nn\\
&&\hspace{.75cm} +\frac{B_z^*B_z}{2}+\frac{\nb\psi^*\cdot\nb\psi}{2}
 -(B^*_z+\mu\Omega)\Ac(\psi^*+\mu v_z)
  \nn 
\\
&&\hspace{.75cm} -(B^*_z+\lambda^{-1}\Omega)\Gc(\psi^*+\lambda^{-1} v_z)
\nn \\
&&\hspace{.75cm}-\rho \Mc(\psi^*+\lambda^{-1}v_z)-\rho \Kc (\psi^*+\mu v_z)\bigg\}=0\,.
 \label{var_pri_main}
\end{eqnarray}
Note, in \eqref{var_pri_main} the Casimir $\Cc_m$  with the  harmonic function $\Phi$ has been omitted. 
 
For the first variation of \eqref{var_pri_main}  to vanish, the coefficients of the arbitrary variations must separately vanish, yielding the following conditions:
\begin{eqnarray}
\delta\rho &:&  \frac{v^2}{2}+\left[\rho U(\rho)\right]_\rho-\Mc(\phi)-\Kc(\varphi)
\nn\\
&-&\frac{d_e^2}{\rho^2}\bigg\{\frac{1}{2}(\Delta \psi)^2+\frac{1}{2}|\nb B_z|^2-\nb B_z\cdot \nb \left[\Ac(\varphi)+\Gc(\phi)\right]\nn \\
&&+\Delta\psi \big[(B_z^*+\mu\Omega)\Ac'(\varphi)+(B_z^*+\lambda^{-1}\Omega)\Gc'(\phi)
\nn\\
&&+\rho(\Mc'(\phi)+\Kc'(\varphi))\big]\bigg\}=0\,,
 \label{bernoulli_sym_1}\\
\delta v_z &:&   \rho v_z- \lambda^{-1}\rho \Mc '(\phi)-\mu \rho \Kc '(\varphi) 
\label{bernoulli_sym_2}\\
&& -\mu(B_z^*+\mu \Omega)\Ac'(\varphi)  -\lambda^{-1}(B_z^*+\lambda^{-1}\Omega)\Gc '(\phi)=0 \,,
\nn \\
\delta\chi &:&  \nb\cdot( \rho \nb \chi)-[\rho,\Upsilon]=\mu \Delta \Ac(\varphi)+\lambda^{-1} \Delta \Gc(\phi) \,,\label{rho_Omega}\\
\delta \Upsilon &:& \hspace{0.1cm} \nb\cdot(\rho\nb\Upsilon)=[\chi,\rho]\,, \label{zeta_eq}\\
\delta B_z^*&:&  B_z=\Ac(\varphi)+ \Gc(\phi) \,,\label{B_z_eq}\\
\delta \psi^* &:& \Delta \psi +\rho \Mc '(\phi)+ \rho \Kc '(\varphi)
\nn\\
&&+(B_z^*+\mu \Omega)\Ac'(\varphi)+(B_z^*+\lambda^{-1} \Omega)\Gc '(\phi)=0 \,,
 \label{psi_gs}
\end{eqnarray}
where $\phi:=\psi^*+\lambda^{-1}v_z$, $\varphi:=\psi^*+\mu v_z$ and $ ' $ denotes the derivative with respect to  argument.  For the derivation of the equilibrium equations above we used the expressions for $B_z^*$, $\psi^*$ in terms of the ordinary magnetic field variables $\psi$ and $B_z$ according to  $\bB^*:=\bB+d_e^2\nb\times (\rho^{-1}\nb\times\bB)=B_z^*(x,y)\hat{z}+\nabla\psi^*(x,y)\times\hat{z}$
with $\bB=B_z(x,y)\hat{z}+\nabla\psi(x,y)\times\hat{z}$:
\begin{eqnarray}
B_z^*&=&B_z-d_e^2\nb\cdot (\rho^{-1}\nb B_z)\,, \label{B_z_star} \\
\psi^*&=&\psi -d_e^2 \rho^{-1}\Delta \psi \,.
\label{psi_star}
\end{eqnarray}

Equation \eqref{bernoulli_sym_1} is a Bernoulli law, which describes the effects of macroscopic ion flows and  electron inertia on  the total pressure. Using  \eqref{B_z_eq} and \eqref{psi_gs}, the  Bernoulli equation takes the form
\begin{eqnarray}
\tilde{P}(\rho)&=&\rho \left[\Mc(\phi)+\Kc(\varphi)\right]-\rho \frac{v^2}{2}
\nn\\
&&\hspace{.75cm}-\frac{d_e^2}{2\rho}\left[(\Delta\psi)^2+|\nb B_z|^2\right]\,,
 \label{bernoulli_gen}
\end{eqnarray}
with the components of the flow velocity being described by the equations \eqref{bernoulli_sym_2}--\eqref{zeta_eq} and $\tilde{P}:=\rho [\rho U(\rho)]_\rho={\gamma} P/({\gamma-1})$, where $P$ is the total pressure (see Sec.~\ref{sec_II}). 
 Also note that Eqs.~\eqref{B_z_eq} and \eqref{B_z_star} can be used to express the quantity $B_z^*$ in terms of the arbitrary functions $\Ac$ and $\Gc$, 
\begin{eqnarray}
B_z^*&=&\Ac(\varphi)+\Gc(\phi)
\label{B_z_A_G}\\
&&\hspace{.5cm}-d_e^2\nb \cdot\left[\rho^{-1}\nb \Ac(\varphi)\right] -d_e^2\nb \cdot\left[\rho^{-1}\nb \Gc(\phi)\right]\,.
\nn
\end{eqnarray}

\subsection{The Grad-Shafranov-Bernoulli system}

We can show (see Appendix B) that  \eqref{bernoulli_sym_2} and \eqref{psi_gs}, with the help of \eqref{rho_Omega},  \eqref{zeta_eq},  and  the definition \eqref{psi_star}, can be written as a Grad-Shafranov-like system of the form
\begin{eqnarray}
&&\alpha_1\Ac'(\varphi)\nb\cdot\left(\frac{\Ac'(\varphi)}{\rho}\nb\varphi\right)+\alpha_2\rho (\varphi-\phi)
 \label{gs_xmhd_1}\\
&&\hspace{1cm}-\alpha_3\frac{\rho}{d_e^2}\left(\psi-\frac{\varphi-\lambda\mu\phi}{1-\lambda\mu}\right)\nn\\ &&\hspace{2cm}
=\left[\Ac(\varphi)+\Gc(\phi)\right]\Ac'(\varphi)+\rho\Kc'(\varphi)\, ,
\nn\\
&&\gamma_1\Gc'(\phi)\nb\cdot\left(\frac{\Gc'(\phi)}{\rho}\nb\phi\right)+\gamma_2\rho (\varphi-\phi) 
 \label{gs_xmhd_2}\\
&&\hspace{1cm}+\gamma_3\frac{\rho}{d_e^2}\left(\psi-\frac{\varphi-\lambda\mu\phi}{1-\lambda\mu}\right)\nn \\ &&\hspace{2cm}
=\left[\Ac(\varphi)+\Gc(\phi)\right]\Gc'(\phi)+\rho\Mc'(\phi)\, ,\nn\\
&&\Delta\psi=\frac{\rho}{d_e^2}\left(\psi-\frac{\varphi-\lambda\mu\phi}{1-\lambda\mu}\right)\,, \label{gs_xmhd_3}
\end{eqnarray}
where
\begin{eqnarray}
\alpha_1&=&\mu^2+d_e^2\,, \quad
\alpha_2=\frac{\lambda^2}{(1-\lambda\mu)^2}\,, \quad \alpha_3=\frac{1}{1-\lambda\mu}\,,
\nn\\
 \gamma_1&=&\lambda^{-2}+d_e^2\,, \quad \gamma_2=-\alpha_2\,, \quad \gamma_3=\lambda\mu\, \alpha_3\,.
\end{eqnarray}
 The above equilibrium equations are coupled to the Bernoulli law \eqref{bernoulli_gen}, comprising  a Grad-Shafranov-Bernoulli (GSB) system. The existence of three coupled equations for three different flux functions, namely $\psi,\phi,\varphi$,  is a direct verification that in the XMHD model the ions and the electrons are allowed to move individually and separate from the magnetic surfaces, forming their own flow surfaces. Upon specifying the free functions $\Ac=\Ac(\varphi)$, $\Gc=\Gc(\phi)$, $\Kc=\Kc(\varphi)$ and $\Mc=\Mc(\phi)$ and adopting an equation of state $P=P(\rho)$,  one can in principle solve the GSB system, at least numerically, to determine the functions $\varphi$, $\phi$, $\psi$ and $\rho$. The level sets of the flux function $\psi$ give the magnetic surfaces on which the magnetic field lines lie. From $\varphi$ and $\phi$ we can compute $\psi^*$ 
 \begin{equation}
 \psi^*=\frac{\varphi-\lambda\mu\phi}{1-\lambda\mu}\,,
  \label{psi_star_varphi_phi}
 \end{equation}
while the poloidal ion flow velocity is given by
\begin{equation}
\bv_p=\frac{1}{\rho}\left(\mu\Ac'(\varphi)\nb\varphi+\lambda^{-1}\Gc'(\phi)\nb\phi\right)\times\hz\,,
 \label{v_p_xmhd}
\end{equation}
 and the longitudinal velocity component  follows from  
\begin{equation}
v_z=\frac{\lambda}{1-\lambda\mu}(\phi-\varphi) \,.
\label{v_z_xmhd}
\end{equation}
Note that the longitudinal component of the magnetic field is directly related to $\varphi$ and $\phi$ through \eqref{B_z_eq} and the poloidal field is simply given by $\bB_p=\nb\psi\times\hz$. Thus all equilibrium quantities of interest can be specified upon solving the system \eqref{gs_xmhd_1}--\eqref{gs_xmhd_3}.
\subsection{Special cases of equilibria}

\subsubsection{Equilibria with longitudinal flow $(\bv_p=0)$}

From \eqref{v_p_xmhd}, requiring $\bv_p=0$ we deduce that
\begin{equation}
\Gc(\phi)=-\lambda\mu\Ac(\varphi) \label{AG}\,,
\end{equation}
hence $\varphi=f(\phi)$. According to \eqref{psi_star_varphi_phi}, $\psi^*=(f(\phi)-\lambda\mu\phi)/(1-\lambda\mu)$ i.e. two sets of flux surfaces exist, the electron surfaces and the magnetic surfaces. The ions and the electrons can flow in the poloidal direction on the same surfaces,  but their relative velocities are constrained so that the  total poloidal velocity vanishes. 
Substituting \eqref{AG} into the system \eqref{gs_xmhd_1}--\eqref{gs_xmhd_2} and using $\varphi=\varphi(\psi^*)$, $\phi=\phi(\psi^*)$ yields 
\begin{eqnarray}
&&\gamma_1\Gc'(\psi^*)\nb\cdot\left(\frac{\Gc'(\psi^*)}{\rho}\nb\psi^*\right)+\gamma_2\rho(\varphi-\phi)
\label{gs_longi_1} \\ 
&&\hspace{1cm}+\gamma_3\frac{\rho}{d_e^2}\left(\psi-\psi^*\right)\nn \\
&&\hspace{2cm}
=\frac{\lambda\mu-1}{\lambda\mu}\Gc(\psi^*)\Gc'(\psi^*)+\rho\Mc'(\psi^*)\,,\nn\\
&&\Delta\psi=\frac{\rho}{d_e^2}\left(\psi-\psi^*\right)\label{gs_longi_2}\, ,
\end{eqnarray} 
 with $\varphi-\phi$ given by 
 \begin{equation}
 \varphi-\phi=(\mu-\lambda^{-1})\left[\mu \Kc'(\psi^*)+\lambda^{-1}\Mc'(\psi^*)\right]\, . \label{f_phi}
 \end{equation}

\subsubsection{Static equilibria}

For the case of static XMHD equilibria, where the macroscopic flow is neglected completely, we require additionally $v_z=0$. Hence the flux functions $\phi$ and $\varphi$ are equal to $\psi^*=\psi -d_e^2 \rho^{-1}\Delta \psi$ i.e. $f(\phi)=\phi=\psi^*$. 
Hence Eqs.~\eqref{gs_longi_1}-\eqref{gs_longi_2} reduce to 
\begin{eqnarray}
&&\gamma_1\Gc'(\psi^*)\nb\cdot\left(\frac{\Gc'(\psi^*)}{\rho}\nb\psi^*\right)
+\gamma_3\frac{\rho}{d_e^2}(\psi-\psi^*) \nn \\ 
&&\hspace{1.5cm}=\frac{\lambda\mu-1}{\lambda\mu}\Gc(\psi^*)\Gc'(\psi^*)+\rho\Mc'(\psi^*)\,,
\label{static_xmhd_1}\\
&&\Delta\psi=\frac{\rho}{d_e^2}(\psi-\psi^*)\, . \label{static_xmhd_2}
\end{eqnarray}
As above, two sets of flux surfaces exist, the electron-ion surfaces and the magnetic surfaces. Note that the electrons and the ions are allowed to move (in order to carry the electric current) but their velocities should satisfy the constraint $m_i\bv_i+m_e\bv_e=0$. In this static case,  the Bernoulli equation \eqref{bernoulli_gen} becomes
\begin{equation}
\tilde{P}=\rho \left[\Mc(\psi^*)+\Kc(\psi^*)\right]-\frac{d_e^2}{2\rho}\left[(\Delta\psi)^2+|\nb B_z|^2\right]\,,
 \label{bernoulli_static}
\end{equation}
closing the GSB system.

 
\subsubsection{Hall MHD equilibria}

The HMHD GSB equilibrium equations can be obtained from  the system of the equations \eqref{gs_xmhd_1}--\eqref{gs_xmhd_2} and \eqref{bernoulli_gen} upon setting  $d_e= 0$. To take properly this limit one should substitute the third term of the lhs of Eqs.~\eqref{gs_xmhd_1} and \eqref{gs_xmhd_2} by \eqref{gs_xmhd_3}. Adopting $(\lambda,\mu)=(\lambda_+,\mu_+)$, for $d_e \rightarrow 0$ we have $\mu\rightarrow 0$ and $\lambda^{-1}\rightarrow d_i$; therefore,  the independent flux functions are the poloidal magnetic flux function $\psi$ and the ion flow function $\phi:=\psi+d_i v_z$. Using the definition of $\phi$,   $v_z$ becomes  
\begin{equation}
v_z=d_i^{-1}(\phi-\psi)\,.
 \label{v_z_hall}
\end{equation} 
Also,  from   \eqref{v_p_xmhd} we take
\begin{equation}
\bv_p=\frac{d_i}{\rho}\Gc'(\phi)\bB_{ip}\,, \qquad \text{where} \qquad \bB_{ip}:=\nb\phi\times\hz\,.
 \label{vp_bip}
\end{equation}
Next, with $d_e=0$  Eqs.~\eqref{gs_xmhd_1}--\eqref{gs_xmhd_2}, in view of \eqref{gs_xmhd_3}, reduce to 
\begin{eqnarray}
&&d_i^2\Gc'(\phi)\nb\left(\frac{\Gc'(\phi)}{\rho}\nb \phi\right)
+\frac{\rho}{d_i^2}(\phi -\psi)
\nn\\
&&\hspace{1cm} -\left[\Gc(\phi)+\Ac(\psi)\right]\Gc'(\phi)-\rho \Mc'(\phi)=0\,, \label{gs_hall_baro_1}\\
&&\Delta\psi+\frac{\rho}{d_i^2}(\phi-\psi) +\rho \Kc'(\psi)
\nn\\
&&\hspace{1cm} +\left[\Gc(\phi)+\Ac(\psi)\right]\Ac'(\psi)=0\,.
 \label{gs_hall_baro_2}
\end{eqnarray}
Finally we close the system by writing the Bernoulli equation \eqref{bernoulli_gen} with $d_e=0$, 
in terms of $\rho$  and the ion and magnetic flux functions. To do so we express the kinetic term using  \eqref{v_z_hall} and \eqref{vp_bip} arriving at 
\begin{eqnarray}
\tilde{P}(\rho)&=&\rho\left[\Kc(\psi)+\Mc(\phi)-\frac{(\phi-\psi)^2}{2d_i^2}\right]
\nn\\
&&\hspace{.75cm} -\frac{d_i^2}{2\rho}\left(\Gc'(\phi)\right)^2|\nabla\phi|^2\,.
 \label{bernoulli_hall_2}
\end{eqnarray}
 To summarize,   translationally symmetric barotropic Hall MHD equilibria are governed by the GSB system \eqref{gs_hall_baro_1}--\eqref{bernoulli_hall_2}
 with $\Kc(\psi)$, $\Mc(\phi)$, $\Ac(\psi)$, $\Gc(\phi)$ being arbitrary functions and the pressure $\tilde{P}(\rho)$ obeys a barotropic equation of state. These are the barotropic translationally symmetric counterparts of the baroclinic axisymmetric equilibrium equations derived by an Euler-Lagrange variational principle in \cite{hame_1} and of the barotropic axisymmetric equilibrium equations derived in \cite{thro_tass} by a direct projection of the 3D equilibrium equations. Other  derivations of the  two-fluid equilibrium equations, which do not ignore electron inertia, have been made by various authors, e.g.\ \cite{lcstein,mcclem,goed}.
 As expected the sets of equilibrium equations derived there, are mostly of the type of the system \eqref{gs_xmhd_1}--\eqref{gs_xmhd_3} because  XMHD is closer to a full two-fluid description than HMHD.

\begin{figure}[h]
\includegraphics[scale=0.65]{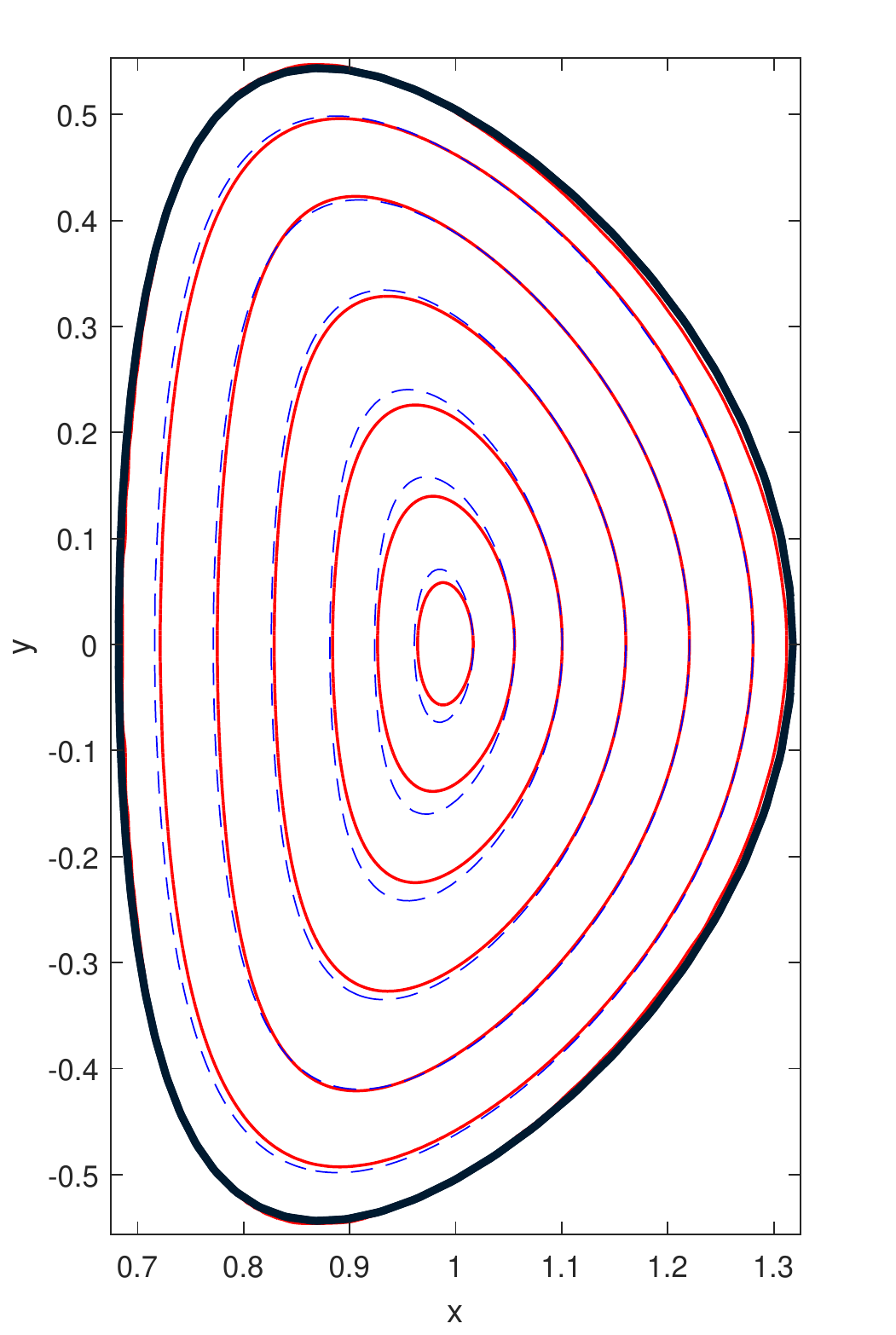} 
\caption{Ion flow surfaces (solid-red) ($\phi=$constant)  and  magnetic surfaces (dashed-blue) ($\psi=$constant) with dimensionless Hall parameter $d_i=0.03$ (normalized ion skin depth) for  a ``straight" Tokamak HMHD equilibrium. The solid black line represents the boundary.  Departure of the flow surfaces from the magnetic surfaces due to the Hall term in Ohm's law is observed, with a  separation distance of the order of $0.04\,L_0$.} \label{figure_1}
\end{figure}

Despite the simpler structure of HMHD, the system of  \eqref{gs_hall_baro_1}--\eqref{bernoulli_hall_2}, forms rather complex classes of equilibria. It requires the simultaneous solution of two coupled nonlinear PDEs, the Grad-Shafranov equations,  which are additionally coupled to a Bernoulli equation and generally the existence of equilibrium solutions is not guaranteed. Due to this strong coupling, studies of two-fluid equilibria have been carried out numerically e.g.\ see \citep{stein_ish_1,guaz_bett}. Here we follow this approach for Hall MHD, giving an example of an equilibrium configuration (Fig.~\ref{figure_1}) computed by means of a simple finite difference iterative code, implemented on Matlab. More information and possible improvements of this computation will be given in a future work. For sake of clarity  we mention that we used an MHD initial guess for $\psi$ and the ion flux function $\phi$ was initialized on the basis of this initial guess. The initial density $\rho$ was set as a linear function of initialized flux function. These quantities are used for the calculation of their updated counterparts in the next iteration and so on until the resulting state converges.  For this particular example we adopted the following choices:
\begin{eqnarray}
\Gc(\phi)&=g_0+g_1\phi +g_2 \phi^2\,,\nn\\
\Ac(\psi)&=a_0+a_1\psi +a_2 \psi^2 \,,\nn\\
\Mc(\phi)&=m_0+m_1\phi+m_2 \phi^2\,, \nn \\
\Kc(\psi)&=k_0+k_1\psi +k_2 \psi^2\,,\nn\\
\tilde{P}(\rho)&=p_1\rho^\gamma\,,  \hspace{1.85cm}& \label{eq_ansatz}
\end{eqnarray}
where $\gamma=5/3$ is the specific heat ratio.
 Note, because we assumed  the plasma is barotropic, $p_1$ can be a constant or at most a function of $\rho$. If we assume additionally  that $p_1$ is a function of $\psi$ and $\phi$,  then the mass density should also be a function of $\psi$ and $\phi$ and,  due to the Bernoulli equation \eqref{bernoulli_hall_2}, $v^2$ should be a function of the poloidal ion and magnetic fluxes, a property  that  demands  certain restrictions on the permissible equilibrium configurations. The present study though, can be  extended to the more generic case of baroclinic closure, i.e., when the internal energy is a function of the density and specific entropies \cite{hame_1}, which yield  dependence of the pressure on the flux functions without restricting the equilibria. This will be considered in our future work.  

For the computation of the equilibrium we imposed Dirichlet boundary conditions on the  fluxes $\psi$ and $\phi$ on a D-Shaped boundary relevant to  fusion experiments with elongation $\kappa=1.7$ and triangularity $\delta=0.4$. In Fig.~\ref{figure_1} we observe the ``departure" of the flow surfaces from magnetic surfaces,  a result  qualitatively consistent with the configurations presented in \cite{guaz_bett} where the baroclinic, axisymmetric, HMHD equilibrium equations were solved by means of the FLOW2 code.  The observed departure is due to the Hall term $d_i{\bJ\times\bB}/{\rho}$ in Ohm's law, which ``breaks" the frozen  flux condition of  ideal MHD. In Hall MHD the flow surfaces are frozen into the ``ion fluid" while the magnetic surfaces are frozen into the ``electron fluid". An estimate of the poloidal separation distance $\Delta r$,   a measure of the departure of the ion flow surfaces from  the magnetic surfaces, was given in  \cite{hame_1}.  For typical Tokamak experiments this quantity is of the order of the ion  poloidal Larmor radius, which is used as a typical step size in neoclassical transport studies. The separation distance can be approximated by
$\Delta r\sim d_i {v_z}/{B_p}$.  For  our computed equilibrium depicted in Fig.~\ref{figure_1},  the normalized poloidal separation distance is $\overline{\Delta r}\sim 0.04$ (for $d_i=0.03$ and using the average values of $v_z$ and $B_p$).

 Although our purpose for this numerical example was to demonstrate the qualitative way  ion surfaces depart  from the magnetic surfaces, which is predicted by the HMHD theory, we mention briefly some equilibrium characteristics of our example. The maximum $\beta$ in the plasma core is $\beta_{max}=1.2\%$, the current density profile is peaked on axis, i.e.,  it appears to have a maximum in the central region with  maximum values of the order of $1\times B_0/(\mu_0L_0)$, while it reverses in the outer region. The plasma response to the external magnetic field is purely diamagnetic, since  the center drops to $0.8\times B_0$ from $1\times B_0$ at the boundary. Lastly, the flow in the $z$-direction is peaked on axis with a maximum value at $0.25\times v_A$, where $v_A=B_0/\sqrt{\mu_0\rho_0}$ and the poloidal velocity component has a maximum value of $0.1\times v_A$. 
  The constants $B_0$, $\rho_0$,  and $L_0$ are reference values for the magnetic field, the mass density,  and the characteristic length scale, respectively. The values and the shapes of the profiles can be adjusted by regulating the free parameters in the Ansatz \eqref{eq_ansatz} and adding some additional nonlinear terms. However, for sake of simplicity, here we consider this Ansatz with parametric values that favor fast convergence and results in  configurations with distinct surface separation.

As a final note, a similar numerical procedure as that employed above for Hall MHD equilibria, can be utilized for the numerical integration of the GS systems \eqref{gs_longi_1}--\eqref{gs_longi_2} and \eqref{static_xmhd_1}--\eqref{static_xmhd_2} for XMHD equilibria with longitudinal flows and static XMHD equilibria respectively. 
%

\section{Conclusion \label{sec_V}}

In this paper we presented  the Hamiltonian formulation of translationally symmetric barotropic extended magnetohydrodynamics. We derived  the symmetric Casimir integrals of motion and produced the Energy-Casimir variational principle for obtaining the generalized equilibrium equations, which govern XMHD stationary states. These states may be particularly interesting for the study of 2D collisionless reconnection configurations. Also,  since two-fluid effects become significant for smaller
length scales, increased values of plasma $\beta$, and flows approaching the ion diamagnetic drift speed,  equilibrium studies based on XMHD equations, could be useful for an adequate description of magnetically confined plasmas with such characteristics.  The equilibrium system of equations were shown to be a Grad-Shafranov-Bernoulli type, and we studied special cases of XMHD equilibria and HMHD equilibria with arbitrary flow.  In the case of HMHD equilibria with arbitrary flow,  we computed  a numerical equilibrium on  a D-Shaped domain, relevant to fusion experiments. The resulting configuration is representative of the predicted separation of the ion-flow and magnetic surfaces. Extension of the present study to  cases of arbitrary symmetry, as done for MHD in  \cite{amp1}, in particular  for helically symmetric configurations,  is in progress and will be published in a future work.  


\section*{Acknowledgements}
This work has been carried out within the framework of the EUROfusion Consortium and has received funding from (a) the National Programme for the Controlled Thermonuclear Fusion, Hellenic Republic and (b) Euratom
research and training program 2014-2018 under Grant Agreement No. 633053. The views and opinions expressed
herein do not necessarily reflect those of the European Commission.  PJM  was supported by  the U.S. Department of Energy Contract DE-FG05-80ET-53088 and  a Forschungspreis from the Alexander von Humboldt Foundation. He  warmly acknowledges the hospitality of the Numerical Plasma Physics Division of  Max Planck IPP,  Garching.   DAK and GNT would like to thank George Poulipoulis and Apostolos Kuiroukidis for useful discussions regarding the construction of the numerical equilibrium.

\numberwithin{equation}{section}

\begin{appendices}
  \renewcommand\thetable{\thesection\arabic{table}}
  \renewcommand\thefigure{\thesection\arabic{figure}}

\section{Derivation of the symmetric Poisson bracket (\ref{ts_poisson})}

\noindent{\it a. Compressional part:}
 \begin{equation}
 \{F,G\}^{c}=-\int_Vd^3x\,\left(F_\rho\nb\cdot G_\bv-G_\rho\nb\cdot F_\bv\right)\,.
 \end{equation}
Using the relation $\nb\cdot F_\bv=-\Delta F_w$ we obtain 
\begin{equation}
\{F,G\}^{c}_{_{TS}}=\int_Dd^2x\,\left(  F_\rho\Delta G_w-G_\rho\Delta F_w\right)\,.
\end{equation}

\medskip
\noindent{\it b. Vortical part:}
\begin{equation}
\{F,G\}^{v}=\int_Vd^3x\, \rho^{-1}\left(\nb\times\bv\right)\cdot\left(F_\bv\times G_\bv\right)\,.
\end{equation}
Using  \eqref{tran_sym_vel_field}  and   $\Omega:=-\Delta\chi$,  the vorticity is
\begin{equation}
\nb\times\bv=\Omega\hz+\nb v_z\times \hz \label{curl_v}\,;
\end{equation}
therefore we have
\begin{eqnarray}
\rho^{-1}\left(\nb\times\bv\right)\cdot\left(F_\bv\times G_\bv\right)&=&\rho^{-1}\Omega\hz\cdot \left(F_\bv\times G_\bv\right)_z
\nn\\
&&\hspace{-.75cm} + \rho^{-1}(\nb v_z\times\hz)\cdot\left(F_\bv\times G_\bv\right)_p\,.
\nn
\end{eqnarray}
The subscripts $z$ and $p$ above denote the $z$ and the poloidal component respectively, which read as follows:
\begin{eqnarray}
\left(F_\bv\times G_\bv\right)_z&=&\nb F_w\times \nb G_w-\nb F_\Omega\cdot \left(\hz\times \nb G_\Omega\right)\hz
\nn\\
&&\hspace{-0.5cm} + (\nb F_w\cdot\nb G_\Omega) \hz- (\nb F_\Omega \cdot \nb G_w) \hz\,, \\
\left(F_\bv\times G_\bv\right)_p&=&F_{v_z}\nb G_\Omega-G_{v_z}\nb F_\Omega
\nn\\
&&\hspace{-0.5cm}+ F_{v_z}\nb G_w\times \hz-G_{v_z}\nb F_w\times\hz\,.
\end{eqnarray}
Using $[a,b]=(\nb a\times \nb b)\cdot\hz$ , 
\begin{eqnarray}
\rho^{-1}\Omega\hz\cdot \left(F_\bv\times G_\bv\right)_z&=&
\rho^{-1}\Omega\big(\left[F_\Omega,G_\Omega\right]
 \label{A_toroidal}\\
&&\hspace{-2.5cm} +\nb F_w\cdot \nb G_\Omega-\nb F_\Omega \cdot\nb F_w+\left[F_w,G_w\right]\big)\,,
\nn \\
\rho^{-1}(\nb v_z\times\hz)\cdot\left(F_\bv\times G_\bv\right)_p&=&
\nn\\
&& \hspace{-2.5cm}\rho^{-1}\big(F_{v_z}[G_\Omega,v_z]-G_{v_z}[F_\Omega,v_z]  \\
&&\hspace{-2.5cm}+F_{v_z}\nb v_z\cdot \nb G_w-G_{v_z}\nb v_z\cdot \nb F_w\big)\,,
\nn
\end{eqnarray}
 integrating over the domain $D$,  and  exploiting   \eqref{identity}, it gives 
\begin{eqnarray}
&&\int_Dd^2x\,\rho^{-1}(\nb v_z\times\hz)\cdot\left(F_\bv\times G_\bv\right)_p
\nn\\
&&\hspace{.5cm} =\int_Dd^2x\, v_z\big\{[F_\Omega,\rho^{-1}G_{v_z}]
 -[G_\Omega,\rho^{-1}F_{v_z}]
\nn \\
&&\hspace{1cm}+\nb F_w\cdot \nb (\rho^{-1}G_{v_z})-\nb G_w\cdot \nb (\rho^{-1}F_{v_z})
\nn\\
&&\hspace{1cm}+\rho^{-1}F_\Upsilon G_{v_z}-\rho^{-1}G_\Upsilon F_{v_z}\big\}\,.
 \label{A_poloidal}
\end{eqnarray}
Integrating \eqref{A_toroidal} over $D$ and adding it to \eqref{A_poloidal} gives  the vortical part of the translationally symmetric bracket.

\medskip
\noindent{\it c. Magnetic field - flow part:} 
The magnetic field-flow (MHD) contribution   is 
\begin{eqnarray}
\{F,G\}^{mf}&=&\int_Vd^3x\,\rho^{-1}\bB^*\cdot\big[F_\bv\times\left(\nb\times G_{\bB^*}\right)
\nn\\
&&\hspace{ 1 cm}-G_\bv\times\left(\nb\times F_{\bB^*}\right)\big]; 
 \label{ff_part}
\end{eqnarray}
hence,  one needs to compute $\rho^{-1}\bB^*\cdot\left[F_\bv\times(\nabla\times G_{\bB^*})\right]$, since the second term of \eqref{ff_part}  follows by interchanging   $F$ and $G$. From \eqref{f_w} and \eqref{curl_f_B}, 
\begin{eqnarray}
&&F_\bv\times\left(\nabla\times G_{\bB^*}\right)=
\nn\\
&&\hspace{.5cm} \left(F_{v_z}\hz+\nb F_\Omega\times\hz-\nb F_w\right)\times \left(G_{\psi^*}\hz+\nb G_{B_z^*}\times\hz\right) \nn \\
&&\hspace{.5cm}=F_{v_z}\nb G_{B_z^*} -G_{\psi^*}\nb F_\Omega + [F_\Omega,G_{B_z^*}]\hz
\nn\\
&&\hspace{1cm} -G_{\psi^*}\nb F_w\times\hz +\nb F_w\cdot \nb G_{B_z}\hz
\end{eqnarray}
and since $\bB^*=B_z^*\hz+\nb\psi^*\times\hz$ one can derive   
\begin{eqnarray}
&&\rho^{-1}\bB^*\cdot F_\bv\times\left(\nabla\times G_{\bB^*}\right)=\rho^{-1}\big \{ B_z^*\big([F_\Omega ,G_{B_z^*}]
\nn\\
&&\hspace{1cm} +\nb F_w\cdot \nb G_{B_z^*}\big) +F_{v_z}[G_{B_z^*},\psi^*] \nn \\
&&\hspace{1cm} + G_{\psi^*}[\psi^*,F_\Omega]-G_{\psi^*}\nb F_w\cdot\nb\psi^*
\big\}\,.
\end{eqnarray}
Integrating over  $D$ and using  \eqref{identity} gives 
\begin{eqnarray}
&&\int_Dd^2x\, \Big\{\rho^{-1}B_z^*\big([F_\Omega,G_{B_z^*}] +\nb F_w\cdot \nb G_{B_z^*}\big) 
\nn\\
&&\hspace{.5cm}   +\psi^*\big([F_\Omega,\rho^{-1}G_{\psi^*}] +[\rho^{-1}F_{v_z},G_{B_z^*}]
\nn\\
&&\hspace{.5cm} 
+\nb F_w\cdot\nb\left(\rho^{-1}G_\psi^*\right)+\rho^{-1}F_\Upsilon G_{\psi^*}\big)\Big\}\,.
 \label{ff_part_term_1}
\end{eqnarray}
The second term of \eqref{ff_part} can be computed by \eqref{ff_part_term_1} upon interchanging   $F$ and $G$.

\medskip
\noindent{\it d. Hall part:}
The Hall part of the bracket \eqref{poisson} is 
\begin{equation}
\{F,G\}^{hall}\!\!= -d_i\!\int_V \!d^3x\, \rho^{-1}\bB^*\cdot\left[\left(\nb\times F_{\bB^*}\right)\times\left(\nb\times G_{\bB^*}\right)\right]\,.
\nn
\end{equation}
Using Eq.~\eqref{curl_f_B} we obtain 
\begin{eqnarray}
\left(\nb\times F_{\bB^*}\right)\times\left(\nb\times G_{\bB^*}\right)&=&[F_{B_z^*},G_{B_z^*}]\hz
\label{hall_inertial_part}\\
&& \hspace{-.75cm}+F_{\psi^*}\nb F_{B_z^*}-G_{\psi^*}\nb F_{B_z^*} \,.
\nn
\end{eqnarray} 
Taking the inner product with $\rho^{-1}\bB^*$ the expression above gives 
\begin{equation}
\rho^{-1}\left( B_z^*[F_{B_z^*},G_{B_z^*}]+F_{\psi^*}[G_{B_z^*},\psi^*]-G_{\psi^*}[F_{B_z^*},\psi^*]\right)\,.
\nn
\end{equation}
which upon integrating over $D$ and using \eqref{identity}  gives  
\begin{eqnarray}
\{F,G\}^{hall}_{TS}&=&-d_i\int_D d^2x\,  \Big\{\rho^{-1}B_z^* [F_{B_z^*},G_{B_z^*}]
\nn\\
&+&\psi^* \left([F_{B_z^*},\rho^{-1}G_{\psi^*}]-[G_{B_z^*},\rho^{-1}F_{\psi^*}]\right)\Big\}\,.
\nn
\end{eqnarray}

\medskip
\noindent{\it e. Electron inertial part:}
\begin{eqnarray}
&&\{F,G\}^{inertial}= 
\nn\\
&&\hspace{.75cm} d_e^2 \int_V d^3x\,\rho^{-1}(\nb\times\bv)\cdot\left[\left(\nb\times F_{\bB^*}\right)\times\left(\nb\times G_{\bB^*}\right)\right]\,.
\nn
\end{eqnarray}
For this  part we   take the inner product of \eqref{hall_inertial_part} with $\rho^{-1}\nabla\times \bv$, where the curl of $\bv$ is given by \eqref{curl_v}. Following the same steps as before, gives 
\begin{eqnarray}
\{F,G\}^{inertial}_{TS}&=&d_e^2\int_Dd^2x\,  \Big\{\rho^{-1}\Omega [F_{B_z^*},G_{B_z^*}]
\nn\\
&+& v_z \left([F_{B_z^*},\rho^{-1}G_{\psi^*}]-[G_{B_z^*},\rho^{-1}F_{\psi^*}]\right) \Big\} \,.
\nn
\end{eqnarray}

\medskip
\section{Derivation of (\ref{gs_xmhd_1})--(\ref{gs_xmhd_3})}

From \eqref{rho_Omega} and \eqref{zeta_eq} we deduce the following (except of the gradients of two harmonic functions
 that  can be neglected): 
\begin{eqnarray}
\mu\nb \Ac+\lambda^{-1}\nb\Gc&=&\rho\nb\chi -\rho \nb \Upsilon \times \hz\,,
 \label{AG_vp} \\
\rho\nb \Upsilon&=&\chi \nb \rho\times \hz\,.
 \label{rho_zeta_eq}
\end{eqnarray}
Taking the cross product of \eqref{AG_vp} with $\hz$ gives
\begin{equation}
\bv_p=\rho^{-1}\left[\mu \Ac'(\varphi)\nb \varphi+\lambda^{-1} \Gc'(\phi)\nb \phi\right]\times\hz \,.
 \label{vp_AG}
\end{equation}
Now from the curl of \eqref{vp_AG} we obtain
\begin{equation}
\Delta\chi=\mu \nb\cdot\left(\frac{\Ac'}{\rho}\nb\varphi\right)+\lambda^{-1}\nb\cdot\left(\frac{\Gc'}{\rho}\nb\phi\right)
=-\Omega \label{delta_chi}\,,
\end{equation}
and substituting the expression above into \eqref{bernoulli_sym_2} gives 
\begin{eqnarray}
&&\left(\mu^2\Ac'+\lambda^{-2}\Gc'\right) 
 \nb\cdot\left(\mu\frac{\Ac'}{\rho}\nb \varphi 
+ \lambda^{-1} \frac{\Gc'}{\rho}\nb \phi\right) =
\label{AG_gs_pre_1}\\
&&\frac{\lambda\rho}{1-\lambda\mu}(\varphi-\phi)+\mu \rho \Kc'
+\lambda^{-1}\rho \Mc'+B_z^*(\mu \Ac'+\lambda^{-1}\Gc')\,,
\nn
\end{eqnarray}
where we used  $v_z= {\lambda}(\varphi-\phi)/({\lambda \mu-1})$, 
which follows from the  definitions of $\varphi$ and $\phi$. Using Eq.~\eqref{B_z_A_G} we can write Eq.~\eqref{AG_gs_pre_1} as
\begin{eqnarray}
&&\left(\mu^3+\mu d_e^2\right)\Ac'\nb\cdot\left(\frac{\Ac'}{\rho}\nb\varphi\right)
\nn\\
&& \hspace{2cm}+\left(\lambda^{-3}+\lambda^{-1}d_e^2\right)\Gc'\nb\cdot\left(\frac{\Gc'}{\rho}\nb\phi\right) \nn \\
&&=\frac{\lambda\rho}{1-\lambda\mu}(\varphi-\phi)+\rho\left(\mu \Kc'+\lambda^{-1}\Mc'\right)
\nn\\
&&\hspace{2.5cm} +\left(\mu\Ac'+\lambda^{-1}\Gc'\right)\left(\Ac+\Gc\right)\,.
 \label{xmhd_gs_1}
\end{eqnarray}
Inserting Eqs.~\eqref{B_z_star} and \eqref{delta_chi} into \eqref{psi_gs} and following a similar procedure as above we derive a second GS-like equation, 
\begin{eqnarray} 
&&\left(\mu^2+d_e^2\right)\Ac'\nb\cdot\left(\frac{\Ac'}{\rho}\nb\varphi\right)+\left(\lambda^{-2}+d_e^2\right)\Gc'\nb\cdot\left(\frac{\Gc'}{\rho}\nb\phi\right) \nn \\
&&\hspace{1cm} =\frac{\rho}{d_e^2}\left(\psi-\frac{\varphi-\lambda\mu\phi}{1-\lambda\mu}\right)+\rho\left(\Kc'+\Mc'\right)
\nn\\
&&\hspace{3cm} +\left(\Ac'+\Gc'\right)\left(\Ac+\Gc\right)\,,
 \label{xmhd_gs_2}
\end{eqnarray}
and $\psi$ is connected to $\varphi$ and $\phi$ by 
\begin{equation}
\Delta\psi=\frac{\rho}{d_e^2}\left(\psi-\frac{\varphi-\lambda\mu\phi}{1-\lambda\mu}\right)
\,. \label{xmhd_gs_3}
\end{equation}
The last equation can be derived from  \eqref{psi_star},  using the definitions of $\varphi$ and $\phi$.  Finally we may refine a bit more the GS system by combining \eqref{xmhd_gs_1} and \eqref{xmhd_gs_2}.  After  careful manipulation,  this system leads to \eqref{gs_xmhd_1}--\eqref{gs_xmhd_2}. 
\end{appendices}

\end{document}